\documentclass[lettersize,journal]{IEEEtran}

\newtheorem{example}{Example}
\newtheorem{thm}{Theorem}
\newtheorem{theorem}[thm]{Theorem}

\newtheorem{remark}{Remark}

\newtheorem{proof}{Proof}
\usepackage{xcolor}
\usepackage{amsmath,amsfonts}
\usepackage{algorithmic}
\usepackage{algorithm}
\usepackage{array}
\usepackage[caption=false,font=normalsize,labelfont=sf,textfont=sf]{subfig}
\usepackage{textcomp}
\usepackage{stfloats}
\usepackage{orcidlink}
\usepackage{url}
\usepackage{soul}
\usepackage[normalem]{ulem}
\usepackage{verbatim}
\usepackage{graphicx}
\usepackage{cite}
\usepackage{todonotes}
\usepackage{tikz}
\usepackage{cleveref}
\usepackage{listings}\usepackage[backgroundcolor=green!30, linecolor=green!60]{mdframed}
\definecolor{uzlmain}{RGB}{0,120,140}
\definecolor{uzlred}{RGB}{200,0,0}
\definecolor{uzlgreen}{RGB}{0,120,140}
\definecolor{defensecolor}{RGB}{0,120,140}
\definecolor{observecolor}{RGB}{200,0,0}
\definecolor{uzlgrey}{RGB}{120,120,120}
\definecolor{diagcolor}{RGB}{255,165,0}

\begin{document}

\title{Towards Efficient Attack Tree Generation from Attack Graphs for Unobservable Target Analysis}

\author{
\IEEEauthorblockN{
Chijioke Eze\,$^{\orcidlink{0000-0002-8545-0160}}$\,,
Aliyu Tanko Ali\,$^{\orcidlink{0000-0003-3841-2054}}$\,,
Leonard Chidiebere Eze\,$^{\orcidlink{0009-0005-0501-1889}}$\,,
Antonello Monti\,$^{\orcidlink{0000-0003-1914-9801}}$\,,
Damas Gruska\,$^{\orcidlink{0000-0002-8517-4688}}$\,, and
Martin Leucker\,$^{\orcidlink{0000-0002-3696-9222}}$\
}

\thanks{This research was sponsored by Slovak Republic under the SRDA grant APVV-23-0292 (DyMAX) and VEGA grant no. 1/0630/25 (eXSec)~\textit{(Corresponding author: Aliyu Tanko Ali)}}

%\thanks{This work was supported by the European Union's EU NextGenerationEU through the Recovery and Resilience Plan for Slovakia under the project No. 09I03-03-V04-00095.~\textit{(Corresponding author: Aliyu Tanko Ali)}}

\thanks{Chijioke Eze and Antonello Monti are with the Institute for Automation of Complex Power Systems, RWTH Aachen University, 52074 Aachen, Germany. Antonello Monti is also with the Fraunhofer Institute for Applied Information Technology, 52068 Aachen, Germany (e-mail:~chijioke.eze@rwth-aachen.de; amonti@eonerc.rwth-aachen.de).}
\thanks{Aliyu Tanko Ali and Martin Leucker are with the Institute for Software Engineering and Programming Languages (ISP), University of Lübeck, Ratzeburger Allee 160, Lübeck, Germany (e-mail:{\{aliyu.ali, leucker\}}@isp.uni-luebeck.de).}
\thanks{Leonard Chidiebere Eze is with the Department of Algebra and Geometry (DAG), Faculty of Mathematics, Physics and Informatics (FMFI),
Comenius University, Mlynska Dolina, 842 48 Bratislava, Slovakia (e-mail: leonard.eze@fmph.uniba.sk).}
\thanks{Damas Gruska is with the Department of Applied Informatics (DAI), Faculty of Mathematics, Physics and Informatics (FMFI),
Comenius University, Mlynska Dolina, 842 48 Bratislava, Slovakia (e-mail: gruska@fmph.uniba.sk).}
}

\markboth{}{}

\maketitle

\begin{abstract}
Attack trees provide a structured approach to security analysis but traditionally require prior knowledge of the attacker's objective. In practice, this objective often remains hidden, even when some attacker actions are partially detected. This paper presents a methodology for generating attack trees without assuming a predefined target by modeling the threat environment as a constraint-weighted graph and applying temporal reachability analysis. The approach identifies probable attacker locations from partial observations and a time budget, carrying traversal cost as an attack-tree attribute. It then extracts target-oriented subgraphs that preserve feasible attack paths while eliminating irrelevant nodes. We demonstrate the framework across three scenarios: a single office building, a multi-building industrial complex, and a real IoT device-communication network recovered from the CICIoT2023 attack dataset. Compared with the baselines, our approach achieves a lower reduction factor while maintaining complete target preservation, substantially outperforming topology-only methods that ignore temporal constraints. We further perform a sensitivity analysis across 66 weight configurations, showing that target identification is stable when strategic value receives at least 20\% of the weight in the single building and at least 50\% in the multi-building complex, while additional experiments reveal sensitivity to perturbations of the underlying strategic-value prior.
\end{abstract}


\section{Introduction}

\IEEEPARstart{A}{ttack} trees (ATs) have long served as a foundational formalism for structured security analysis, enabling defenders to decompose high-level adversarial goals into concrete, actionable attack steps. Their effectiveness, however, rests on the critical assumption that the attacker’s objective is known in advance. In real-world cyber-physical and Internet of Things (IoT) environments, this assumption is frequently violated. Intrusions are often detected only through partial observations, such as a sensor anomaly, an unauthorized entry captured on camera, or a network flow deviation. Meanwhile, the attacker’s subsequent movements and ultimate target remain hidden. Existing IoT intrusion detection approaches, including rule-based methods and traffic- or protocol-level classification, are largely designed to flag individual malicious events or stages, such as the lateral movement stage of an advanced persistent threat (APT) attack carried over server message block (SMB) or Remote Desktop Protocol (RDP) traffic~\cite{he2023comprehensive}, rather than to reason over the space of plausible multi-step paths an attacker may take afterward. Such methods can confirm that a single step of an attack has occurred, but they offer no mechanism for inferring which downstream locations or assets remain feasible targets once subsequent attacker movement becomes unobservable. Consequently, even when an intrusion is successfully detected, IoT environments often yield only fragmented observations of it, because devices are numerous and heterogeneous, communication is intermittent, and monitoring coverage is inherently sparse.

The problem of \emph{unobservable target analysis}, that is, inferring probable attack targets from limited, noisy observations, has remained an open challenge in threat modeling. Without knowing an attacker's ultimate goal, AT synthesis is not feasible in such threat environments, since constructing an AT requires the root node (which is the attack target) to be specified beforehand. In our prior work presented at NordSec~2025~\cite{ali2025graph}, we introduced a graph-based methodology to address this gap. By modeling the threat environment (including asset states) as a constraint-weighted graph and applying temporal reachability analysis, we reduced the space of possibilities to only those locations where the attacker could be present. We then ranked high-value targets within this space using a composite strategic metric and extracted a target-oriented subgraph from which a minimal AT can be constructed. This approach provided a formal pathway from partial observations to synthesizing ATs without presupposing the attacker’s goal.

While the earlier work established the theoretical framework, several practical questions remain. For example, how does the method scale to larger, more complex environments such as a multi-building industrial facility or a real IoT device network? How sensitive are the target rankings to the choice of weighting parameters? How does our graph-reduction approach perform against baseline methods such as $k$-hop neighborhood analysis~\cite{nikolentzos2020k}?

In this paper, we aim to answer these questions through \emph{implementation and evaluation}. Specifically, we implement a framework that models the environment, infers likely attacker locations from partial observations, and constructs a minimal AT accordingly. We demonstrate this through (1) a simple three-floor building (replicating the NordSec paper’s working example), (2) a multi-building industrial facility with interconnected structures, and (3) a real IoT device-communication network recovered from the CICIoT2023 dataset~\cite{neto2023ciciot2023}. The scenarios are assessed against \emph{reduction} and \emph{target-ranking} baselines using metrics such as reduction factor, target preservation, and ranking accuracy.

This work extends the previous work~\cite{ali2025graph} with the following contributions:

\begin{itemize}
    \item An extended literature review that covers attack-tree formalisms, IoT threat modeling, intrusion detection, and graph-reduction techniques.
    \item An implementation and evaluation of the framework across three use cases: a single building, a multi-building industrial facility, and a real IoT device-communication network recovered from CICIoT2023.
    \item A comparative analysis against reduction baselines ($k$-hop, forward-only, and all-paths reachability) and target-ranking baselines (degree centrality, value-only, PageRank, and random selection).
    \item A sensitivity study of the strategic ranking function, providing guidance for parameter selection in real-world deployments.
\end{itemize}

The remainder of this paper is organized as follows. Section~\ref{sec:related} reviews related work in ATs, IoT security modeling, and graph reduction. Section~\ref{sec:methodology} presents our methodology. Section~\ref{subsec:temporal-reachability} presents temporal reachability analysis, while Section~\ref{sec:five} demonstrates the use cases and their evaluation. In Section~\ref{sec:six}, we present the results of the analysis. Finally, Section~\ref{sec:conclusion} concludes the paper.

 \section{Related Work}
\label{sec:related}

Our work draws on three research streams: attack-tree formalisms and their use in IoT threat modeling, intrusion detection and attacker-target prediction, and graph-reduction techniques for security analysis.

\subsection{Attack Trees and Graphs in IoT Environments}

ATs were introduced by Schneier as a hierarchical decomposition of adversarial goals~\cite{schneier1999attack}. Since then, numerous extensions~\cite{kordy2014attack} have been proposed that incorporate attributes such as cost, time~\cite{ali2024attack}, and probability~\cite{kordy2016probabilistic}. Automated generation of ATs from system models has been explored by Kordy \textit{et al.}~\cite{kordy2013adtool}, who developed the ADTool for graphical editing and analysis. More recently, Konsta \textit{et al.}~\cite{konsta2024survey} provided a survey of automated AT and attack graph generation, highlighting the gap between model-based synthesis and real-world observability. Model-based synthesis follows two approaches: offline and online. In offline synthesis, ATs are generated from documented data or expert brainstorming~\cite{konsta2024attack,de2024automated,ali2021attack}. Online synthesis involves automatic generation of ATs from detected attack patterns~\cite{ga2003generating}. Although offline model-based synthesis has been widely studied, online model-based synthesis has received little attention.

Despite advances in automated synthesis, generating comprehensive ATs remains labor-intensive, particularly in the context of the IoT. IoT environments introduce challenges that complicate traditional modeling, such as substantial device heterogeneity, dynamic network topologies, and resource constraints. To address these, several works have adapted ATs specifically for IoT threat modeling. For instance, the work in~\cite{sequeiros2024approach} introduced the ATIoT tool, which automates the creation of ATs directly from system descriptions to simplify the modeling of complex IoT architectures. Jawad \textit{et al.}~\cite{jawad2025stridesea} proposed STRIDE-SEA, extending the framework with sensor-specific threats, but their evaluation assumes complete visibility into device states. These works often rely on the assumption of pre-defined attack objectives, an assumption that is increasingly difficult to justify in evolving IoT environments. Security monitoring frequently reveals only partial or initial indicators of compromise, while subsequent attacker behavior remains obscured due to limited observability. As a result, defenders can detect early-stage intrusions but lack the means to reliably infer the ultimate objectives of the attacker.

\subsection{Intrusion Detection and Target Prediction}
Intrusion detection is the process of identifying malicious or unauthorized activities from observed system or network behavior. Beyond detecting ongoing intrusions, target prediction seeks to infer an attacker's likely target or intended objective from available observations, often under uncertainty. Predicting attacker intent under uncertainty has been studied in both cyber and physical security domains~\cite{ali2022states}. In the context of ATs, several works have focused on enriching AT models with quantitative and stochastic attributes to support predictive analysis. For instance, Jhawar \textit{et al.}~\cite{jhawar2016stochastic} introduced a stochastic framework for evaluating attack-defense trees, enabling probabilistic reasoning about attacker success. Later, they proposed a semi-automated method to augment ATs using an annotated library, improving the coverage of known attack patterns~\cite{jhawar2018semi}. Similarly, Gadyatskaya \textit{et al.}~\cite{gadyatskaya2017refinement} presented a refinement-aware generation technique that preserves semantic consistency during tree expansion.

Beyond tree-based models, machine learning and data mining techniques have also been applied to study intrusion detection. Raakesh \textit{et al.}~\cite{raakesh2022machine} employed protocol-based prediction to improve the accuracy of intrusion analysis in IoT devices. More recently, Rebhi \textit{et al.}~\cite{rebhi2025real} used convolutional neural networks (CNNs) and XGBoost for binary classification of network traffic. While effective for detecting low-level indicators at the network or host level, such as anomalous traffic patterns or unauthorized system events, these data-driven methods do not directly model an adversary's strategic goals (high-level adversarial objectives) or the coordinated sequences of physical actions (multi-step physical movements) required to achieve them.

Forecasting attacker behavior in cyber-physical systems remains challenging. Kaiser \textit{et al.}~\cite{kaiser2021attack} surveyed attack forecast methodologies but noted a lack of integrated models that combine cyber observables with physical movement constraints. Early work by Ray and Poolsapassit~\cite{ray2005using} used ATs to identify malicious insider threats, but assumed full observability of user actions. The work in~\cite{garg2018empirical} proposed a methodology for prioritizing attack paths in multi-step attack scenarios by modifying CVSS v3 metrics to compute per-node risk scores, then applying a PageRank model, which iteratively computes the importance of each node based on the number and weight of its incoming connections in the attack graph, to rank complete attack paths. Their work highlights that individual vulnerability scoring is insufficient when attackers chain exploits across multiple hosts, motivating the need for path-level analysis such as the graph reduction approach proposed here.

Our use of temporal reachability is inspired by time-constrained path analysis in transportation networks~\cite{delling2009time}. We adapt this concept to security by incorporating traversal costs, strategic asset valuation, and partial observability, enabling inference of high-value targets from limited observations. Unlike prior machine-learning-based detection systems, our approach explicitly reasons over the space of possible attacker locations and goals. In this setting, attacker locations correspond to system states representing the set of assets that may have been reached by the attacker, while attacker goals represent the target assets or security objectives associated with successful attack execution.

\subsection{Graph Reduction Techniques}
Graph reduction techniques originate in graph theory and algorithm design, where the primary objective is to simplify large graphs while preserving properties relevant to a specific computational task. Examples of graph-reduction techniques include vertex contraction in the Stoer--Wagner minimum-cut algorithm~\cite{stoer1997simple}, vertex and edge deletions used in generating sets of $(k,g)$-graphs \cite{eze2026theoretical}, and sparsification techniques that approximate large networks with fewer edges while maintaining cut or spectral properties~\cite{spielman2008graph}. In theoretical computer science, graph reduction has also been applied to canonical problems, such as graph isomorphism testing (e.g., Weisfeiler-Leman refinement~\cite{weisfeiler1968reduction}) and to structural results, including the graph removal lemma. 

As graph-based models became central in applied domains, particularly information technology and systems engineering, graph reduction techniques evolved into graph summarization, coarsening, and abstraction techniques aimed at enabling scalable reasoning over real-world networks. These techniques group vertices, delete vertices and edges, collapse paths, or eliminate unwanted substructures while retaining reachability and connectivity semantics. In cyber and physical security, graph reduction is required for tractable analysis, particularly for attack graphs~\cite{sheyner2002automated}, which describe all feasible adversarial paths through a system. Although expressive, attack graphs grow exponentially with system size and quickly become intractable. Mell and Harang~\cite{mell2015minimizing} showed that different attack-graph representations can differ by orders of magnitude in size, underscoring the need for reduction techniques that preserve attack semantics while minimizing structural complexity. Recent work continues this direction: Bai \textit{et al.}~\cite{bai2025network} reduce enterprise attack graphs through similarity-based node aggregation and importance-based edge pruning, and Viticchié \textit{et al.}~\cite{viticchie2025advanced} prune 81--98\% of redundant nodes in operational-technology attack graphs while preserving critical paths. These methods reduce by structural importance or path membership.

In the context of intrusion detection and unobservable-target analysis, however, none of these methods incorporate temporal feasibility or the inference of an unobservable target. This matters because existing attack-tree synthesis carries a prerequisite that fails under partial observability: it requires the root (the attack target) to be known in advance. Building on classical reduction principles, our method removes this prerequisite by inferring likely targets from observations and adding temporal feasibility, retaining only nodes on time-bounded paths to the ranked target. The result is a \emph{target-oriented subgraph} that restricts attention to temporally feasible regions of the attack graph and eliminates nodes and edges irrelevant to candidate targets, serving as the intermediate structure from which a minimal AT preserving all feasible attack paths is synthesized.
 \section{Methodological Framework}
\label{sec:methodology}
This section presents the methodological framework with graph-theoretic foundations used throughout this paper. We first formalize the (threat) environment as a weighted graph whose vertices are classified into two categories, namely observable and unobservable nodes. Then we introduce the concept of an AT, a threat modeling tool used to study the various ways an attacker can compromise an asset.

\sloppy
\subsection{Environment Graph Model}

We model the target environment as a directed, weighted graph where nodes represent distinct locations or states and edges represent feasible transitions. We incorporate traversal constraints and a partition of nodes based on observability from a given location. Formally, we define such a graph as follows: 

A \emph{graph} representing the environment is defined as a tuple
\[
G = (V, E, \lambda, \Omega_{t}^{c}),
\]
where each parameter is defined as follows: $V$ is a finite set of nodes representing distinct locations or system states. $E \subseteq V \times V$ is the set of directed edges, where each ordered pair $(u,v) \in E$ represents a directed feasible connection from location $u$ to location $v$. An edge $(u,v)$ exists if an attacker can traverse from $u$ to $v$ without violating structural constraints. $\lambda: V \rightarrow \mathcal{L}$ is a labeling function assigning semantic labels 
from the attribute set $\mathcal{L} = 2^{\mathcal{A}}$, where 
$\mathcal{A} = $ \{\texttt{entry}, \texttt{critical}, \texttt{monitored}, \texttt{target}, \texttt{restricted}, \texttt{classified}, \ldots\}. To capture traversal constraints, we define a weight function
\[
\Omega_{t}^{c}: E \rightarrow \mathbb{R}_{+} \times \mathbb{R}_{+},
\]
where $\Omega_{t}^{c}(u,v) = (t(u,v), c(u,v))$ assigns to each edge $(u,v) \in E$ a traversal time $t(u,v)$ from $u$ to $v$ and a generalized cost $c(u,v)$ from $u$ to $v$. In general $t(u,v) \neq t(v,u)$ and $c(u,v) \neq c(v,u)$. The cost represents physical effort, detection risk, or credential requirements. For convenience, we will always use $t$ for $t(u,v)$ and $c$ for $c(u,v)$. Furthermore, $\Delta t$ denotes elapsed time; that is, $\Delta t = t_{\text{now}}-t_0$. A \emph{path} from a node $v_0 \in V$ to a node $v_k \in V$ is a finite sequence
\[
p = (v_0, v_1, \ldots, v_k)
\]
such that $(v_{i-1}, v_i) \in E$ for all $1 \le i \le k$. Traversal time and generalized cost are dimensionally distinct quantities, so we do not combine them into a single scalar. The \emph{length} of the path $p = (v_0, \ldots, v_k)$ is instead vector-valued,
\[
\begin{gathered}
l(p) = \bigl(t_p,\, c_p\bigr), \quad \text{where}\\[2pt]
t_p = \sum_{i=1}^{k} t(v_{i-1}, v_i), \quad
c_p = \sum_{i=1}^{k} c(v_{i-1}, v_i),
\end{gathered}
\]
where each sum is taken over the directed edges $(v_{i-1}, v_i) \in E$ traversed by $p$. Feasibility is correspondingly governed by a joint budget rather than a single threshold: given $(T, C) \in \mathbb{R}_{+} \times \mathbb{R}_{+}$, a path $p$ is \emph{feasible} under the budget $(T,C)$ if
\[
t_p \leq T \quad \text{and} \quad c_p \leq C .
\]

Keeping the two constraints structurally separate ensures that the model remains dimensionally well defined. None of the scenarios used in our experiment provides independently measured action costs. Edge costs $c(u,v)$ are therefore defender-assigned, dimensionless modeling priors on a normalized scale, as are traversal times and strategic values; Section~\ref{sec:five} gives the values used in each use case. In all experiments, the cost constraint is not enforced because the available data do not provide independently measured action costs. Consequently, no path is excluded on the basis of cost, and the time budget is the only feasibility criterion. Cost is instead carried as an attribute of the synthesized AT: each step is annotated with its edge cost, and the path cost $c_p$ of a branch is the sum of its step costs. This treatment is consistent with our prior work~\cite{ali2025graph}.

A critical aspect of our model is the distinction between locations where adversarial presence can be detected and those where it cannot. We partition the set of nodes $V$ into \textit{observable} (denoted as $V_{\text{obs}})$ and \textit{unobservable} (denoted as $V_{\text{unobs}})$ subsets:
\[
V = V_{\text{obs}} \cup V_{\text{unobs}}, \quad V_{\text{obs}} \cap V_{\text{unobs}} = \emptyset.
\]

For example, a node $v \in V_{\text{obs}}$ is observable if the system provides sufficient information to uniquely determine whether a malicious action is happening at $v$ at a given time. In contrast, a node $v \in V_{\text{unobs}}$ is unobservable if occupancy of $v$ cannot be inferred from available observations.

Observations may change at run time, i.e., what is observable may become unobservable due to sensor failures or loss of sight. We define this observability constraint as a function $\mathcal{O}: V \times \mathbb{R}_{+} \rightarrow \{1,0\}$ that indicates whether a node $v \in V$ is under observation at time $t \in \mathbb{R}_{+}$.
%\[
%\mathcal{O}(v, t) = \begin{cases} \text{1} & \text{if } v \in V_{\text{obs}}  \\ \text{0} & \text{if } v \in %V_{\text{unobs}} 
%\end{cases} 
%\]
\[
\mathcal{O}(v, t) = \begin{cases} \text{1} & \text{if $v$ is under observation at time $t$,} \\ \text{0} & \text{otherwise.}
\end{cases}
\]
Thus $\mathcal{O}(v, t) = 0$ for all $t$ if $v \in V_{\text{unobs}}$, whereas a node $v \in V_{\text{obs}}$ can become temporarily unobserved, as illustrated in Example~\ref{ex:dynamic-observability}.
\begin{example}
\label{ex:dynamic-observability}
Consider a node $u$ that is initially observable at time $t_0$, i.e., $u \in V_{\text{obs}}$ and $\mathcal{O}(u, t_0) = 1$ at time $t_0$. At this time, malicious actions at $u$ can be detected through a camera or other monitoring technology. However, at time $t_1 > t_0$, the monitoring mechanism fails (e.g., due to power loss, communication disruption, or system compromise). Consequently, $\mathcal{O}(u, t_1)$ transitions from $1$ to $0$, rendering $u$ effectively unobservable despite remaining in $V_{\text{obs}}$. From time $t_1$ onward, malicious actions at the location $u$ can no longer be reliably detected. 
\end{example}

\subsection{Attack Tree Formalism}
We model the different ways an attacker could achieve their objectives using an AT formalism. The root of the AT represents the attacker’s ultimate goal of reaching the target node (denoted as $v_{\text{target}}$), while internal nodes represent intermediate goals such as accessing restricted areas or disabling IoT sensors. Leaf nodes correspond to atomic actions or points at which the attacker may be detected, including physical movement, device manipulation, or cyber-physical attacks against IoT components. AND and OR nodes capture conjunctive and alternative strategies, respectively.

\begin{figure}[!ht]
    \centering
    \resizebox{1.0\columnwidth}{!}{\begin{tikzpicture}[
  level 1/.style={sibling distance=8cm, level distance=1.8cm},
  level 2/.style={sibling distance=4cm, level distance=1.8cm},
  every node/.style={rectangle, draw=uzlmain, thick, fill=uzlgreen!10, 
                     text width=3.5cm, align=center, minimum height=0.8cm,
                     font=\small},
  edge from parent/.style={draw=uzlmain!70, thick, -}
]
\node {\textbf{Root Node}\\ \textit{Attacker's Objective}}
  child {
    node {\textbf{Sub-goal 1}\\ \textit{OR}}
    child {node[fill=uzlmain!20] {\textbf{Atomic Attack 1.1}\\ \textit{(Leaf)}}}
    child {node[fill=uzlmain!20] {\textbf{Atomic Attack 1.2}\\ \textit{(Leaf)}}}
  }
  child {
    node {\textbf{Sub-goal 2}\\ \textit{AND}}
    child {node[fill=uzlmain!20] {\textbf{Atomic Attack 2.1}\\ \textit{(Leaf)}}}
    child {node[fill=uzlmain!20] {\textbf{Atomic Attack 2.2}\\ \textit{(Leaf)}}}
  };
\end{tikzpicture}}
    \caption{The Structure of an attack tree}
    \label{fig:at}
\end{figure}
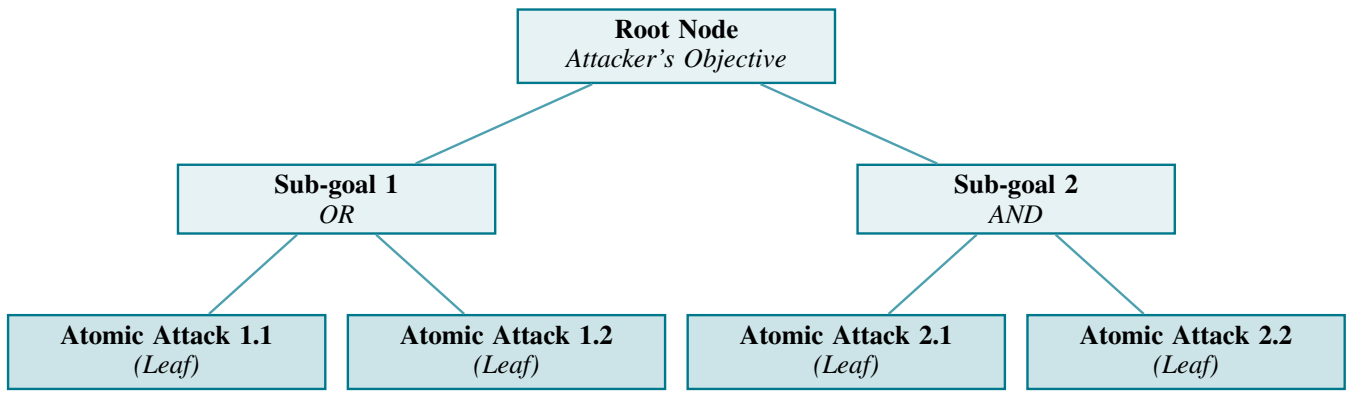

Fig.~\ref{fig:at} shows the structure of an AT. The root node represents the attack objective, and sub-goals 1 and 2 decompose it into alternative ways of achieving it. When a sub-goal cannot be decomposed further, it is represented as an atomic attack action in a leaf node.

In the context of movement through an environment graph, the root node represents the target of the attack, OR nodes represent alternative paths leading to the same location, and AND nodes represent attack steps that must all be completed to achieve the parent node's objective. When multiple actions must be performed in a specific order, we use a sequential-AND (SAND) node, which requires its children to be satisfied in the specified sequence~\cite{jhawar2015attack}.

 \section{Temporal Reachability Analysis}
\label{subsec:temporal-reachability}
Both temporal and operational factors restrict the traversal between nodes. For each edge $(u,v) \in E$, the traversal time $t$ captures the physical distance and environmental conditions, while the cost $c$ reflects detection risk or resource use. Higher costs may be associated with traversing monitored areas or bypassing IoT-controlled barriers, whereas exploiting compromised devices may reduce traversal cost.

In this section, we show how to evaluate whether an attacker can reach a designated target, defined as a protected asset or security objective that constitutes the intended outcome of the attack, within operational time limits. Our goal is to determine not only whether the adversary can reach a protected target within a maximum allowable time, but also which intermediate locations remain viable as the attack progresses. By combining forward and backward reachability analysis, we characterize the attacker’s feasible movement space, identify positions that retain strategic relevance under time pressure, and enable subsequent inference of likely targets, prioritization of critical assets, and construction of the reduced subgraph.

\subsection{Attacker Progress and Temporal Feasibility}

Since edge weights may be asymmetric, the forward and backward reachable sets are generally not symmetric, reflecting direction-dependent traversal costs. Temporal reachability analysis determines whether an attacker can reach the target location within a joint budget $(T, C)$ while satisfying traversal constraints and observability conditions.

\paragraph{Forward Reachable Set}
\label{subsubsec:forward-reachable}

The forward reachable set $\mathcal{R}^{+}(v_0, T, C)$ consists of all nodes reachable from an initial node $v_0$ by a path that is feasible under the budget $(T, C)$. This set captures the attacker’s possible positions as time evolves, accounting for traversal delays and IoT interactions.

\paragraph{Backward Reachable Set}

The backward reachable set $\mathcal{R}^{-}(v_{\text{target}}, T, C)$ consists of all nodes \emph{from which} the candidate target node $v_{\text{target}}$ is reachable by a path feasible under $(T, C)$. This analysis identifies critical locations and IoT assets whose protection can substantially reduce attack feasibility.

\paragraph{Possible Attack Locations}

Let $U \subset V_{\text{obs}}$ denote the set of observed attacker positions, and let $\Delta t$ be the time elapsed since the attacker was last observed, so that $T - \Delta t$ is the remaining time available to reach the target. A node $v \in V$ is a \emph{possible attack location} if it is simultaneously forward-reachable from some observed position within the elapsed time and backward-reachable from the candidate target within the remaining budget. Formally,
\[
\mathcal{A}(T, C)
=
\bigcup_{\substack{u \in U \\ \gamma \in [0, C]}}
\bigl(
\mathcal{R}^{+}(u, \Delta t, \gamma)
\cap
\mathcal{R}^{-}(v_{\text{target}}, T - \Delta t, C - \gamma)
\bigr),
\]
where $\gamma$ denotes the portion of the cost budget expended before reaching $v$. A node $v \in \mathcal{A}(T, C)$ therefore satisfies three explicit conditions: (i) $v$ is reachable from some observed node $u \in U$ within the elapsed time $\Delta t$ and cost $\gamma$; (ii) $v$ can still reach the target within the remaining time $T - \Delta t$ and remaining cost $C - \gamma$; and (iii) the concatenation of the two witness paths forms a single path $u \leadsto v \leadsto v_{\text{target}}$ whose total time and cost respect $T$ and $C$. Because the defining predicate constrains $v$ itself, rather than only $u$ and $v_{\text{target}}$, membership depends on the candidate node.

\begin{theorem}[Soundness]
Let $(T, C) \in \mathbb{R}_{+} \times \mathbb{R}_{+}$. If $v \in \mathcal{A}(T, C)$, then there exist an observed node $u \in U$ and a path $p = u \leadsto v \leadsto v_{\text{target}}$ with
\[
t_p \leq T \qquad \text{and} \qquad c_p \leq C,
\]
that is, $v$ lies on a path from an observed position to the target that is feasible under the budget $(T, C)$.
\end{theorem}

\begin{proof}
Suppose $v \in \mathcal{A}(T, C)$. By definition there exist $u \in U$ and $\gamma \in [0, C]$ with
\[
v \in \mathcal{R}^{+}(u, \Delta t, \gamma) \ \cap \ \mathcal{R}^{-}(v_{\text{target}}, T - \Delta t, C - \gamma).
\]
From $v \in \mathcal{R}^{+}(u, \Delta t, \gamma)$ there is a path $p_1 = u \leadsto v$ with $t_{p_1} \leq \Delta t$ and $c_{p_1} \leq \gamma$. From $v \in \mathcal{R}^{-}(v_{\text{target}}, T - \Delta t, C - \gamma)$ there is a path $p_2 = v \leadsto v_{\text{target}}$ with $t_{p_2} \leq T - \Delta t$ and $c_{p_2} \leq C - \gamma$. Concatenating $p_1$ and $p_2$ yields $p = u \leadsto v \leadsto v_{\text{target}}$ with
\[
t_p = t_{p_1} + t_{p_2} \leq \Delta t + (T - \Delta t) = T,
\]
\[
c_p = c_{p_1} + c_{p_2} \leq \gamma + (C - \gamma) = C .
\]
Hence $p$ passes through $v$ and is feasible under $(T, C)$, as required.
\end{proof}

Furthermore, we show that an increase in the time allowance, with the cost budget held fixed, can only expand the set of possible attack locations.

\begin{theorem}[Monotonicity]
Let $C \in \mathbb{R}_{+}$ be fixed and let $T_1, T_2$ be time bounds. If $T_1 \leq T_2$, then
\[
\mathcal{A}(T_1, C) \subseteq \mathcal{A}(T_2, C).
\]
\end{theorem}

\begin{proof}
Let $v \in \mathcal{A}(T_1, C)$. By definition there exist $u \in U$ and $\gamma \in [0, C]$ with $v \in \mathcal{R}^{+}(u, \Delta t, \gamma)$ and $v \in \mathcal{R}^{-}(v_{\text{target}}, T_1 - \Delta t, C - \gamma)$. Since $T_1 \leq T_2$, we have $T_1 - \Delta t \leq T_2 - \Delta t$, and a path feasible within the smaller remaining budget is also feasible within the larger one, so $v \in \mathcal{R}^{-}(v_{\text{target}}, T_2 - \Delta t, C - \gamma)$. The forward condition and the cost split $\gamma$ are unchanged. Hence $v \in \mathcal{A}(T_2, C)$, and since $v$ was arbitrary, $\mathcal{A}(T_1, C) \subseteq \mathcal{A}(T_2, C)$.
\end{proof}

\begin{remark}
Although feasibility is defined with respect to the joint budget $(T, C)$, the time bound $T$ is the parameter that governs attacker localization in this work. The cost component is retained in the edge-weight function $\Omega_{t}^{c}(u,v) = (t(u,v), c(u,v))$ so that the formulation accommodates multi-constraint feasibility and keeps time and cost dimensionally separate. In all analyses, implementations, and examples reported in this paper, however, the cost constraint is not enforced: path cost is carried as an AT attribute, but no path is excluded on the basis of cost, so $T$ alone determines $\mathcal{A}(T, C)$. This keeps the computation within the scalar shortest-path setting and avoids full bicriteria optimization, while preserving the generality of the bivariate model.
\end{remark}

\subsection{Candidate Target Inference and Strategic Ranking}
In this section, we formalize how potential attacker targets are inferred and prioritized using quantitative metrics derived from the environment graph. Each location is evaluated according to its accessibility, structural influence, and operational importance. Given an inferred attacker entry node \(v_0 \in V\), we rank candidate targets $v_{\text{target}} \in V$  using the following metrics:

\paragraph{Degree Centrality}
The degree centrality measures how well connected a node is within the environment graph; well-connected nodes can reach, or be reached from, many others. For a directed graph we use the total (in- plus out-) degree, normalized to $[0,1]$:
\[
C_{\text{D}}(v) = \frac{\deg^{-}(v) + \deg^{+}(v)}{2(n - 1)},
\]
where $\deg^{-}$ and $\deg^{+}$ are the in- and out-degree and $n=|V|$.

\paragraph{Closeness Centrality}
The closeness centrality measures how \emph{easily a node is reached} from the rest of the graph, which is the pertinent notion for a candidate target. Using the time-weighted shortest paths \emph{into} $v$, we adopt the normalized (Wasserman--Faust~\cite{wasserman1994social}) form
\[
C_{\text{C}}(v) = \frac{k-1}{\sum_{u \rightsquigarrow v} d(u,v)} \cdot \frac{k-1}{n - 1},
\]
where $k$ is the number of nodes that can reach $v$ and $d(u,v)$ is the shortest traversal time from $u$ to $v$. A target reachable quickly from many entry points scores higher; a peripheral sink scores lower.

\paragraph{Strategic Value}
A normalized strategic value encodes operational criticality
\[
S: V \rightarrow [0,1],
\]
which reflects the importance of assets hosted at node \(v\), such as IoT controllers or access-management systems. We treat $S(v)$ as a defender-supplied input rather than a quantity derived by the framework itself. Organizations typically already maintain asset-criticality information, such as risk registers, business-impact assessments, or data-classification policies, that records which systems and locations are most sensitive to compromise; $S(v)$ is the interface through which this existing knowledge enters the ranking function. Any monotonic scoring scheme that reflects relative asset criticality on $[0,1]$ is compatible with the framework, and we do not prescribe how a defender should construct it. This mirrors our treatment of the labeling function $\lambda$ and the observability partition $V_{\text{obs}}/V_{\text{unobs}}$, which are likewise defender-specified rather than derived.

\paragraph{Combined Ranking Function}
These metrics are aggregated into a unified ranking function
\[
R(v) = w_{\text{D}}\, C_{\text{D}}(v) + w_{\text{C}}\, C_{\text{C}}(v) + w_{\text{S}}\, S(v),
\]
where $w_{\text{D}}, w_{\text{C}}, w_{\text{S}} \ge 0$ with $w_{\text{D}}+w_{\text{C}}+w_{\text{S}}=1$ are defender-defined weights. Throughout, the symbol $\alpha$ denotes only the reduction factor (Section~\ref{subsec:temporal-reachability}); the ranking weights are written $w_{(\cdot)}$ to avoid ambiguity.

\begin{table}[h]
\caption{Illustrative candidate ranking with weights $(w_{\text{D}},w_{\text{C}},w_{\text{S}})=(0.2,0.3,0.5)$.}
\label{tab: table 2}
\begin{tabular}{l c c c c c}
\hline
Node & $C_{\text{D}}(v)$ & $C_{\text{C}}(v)$ & $S(v)$ & $R(v)$ \\
\hline
$v_1$ & 0.9 & 0.2 & 0.1 & 0.29 \\
$v_2$ & 0.4 & 0.8 & 0.3 & 0.47 \\
$v_3$ & 0.1 & 0.3 & 1.0 & 0.61 \\
\hline
\end{tabular}
\end{table}

\begin{example}
    Table \ref{tab: table 2} illustrates how $R(v)$ combines degree centrality, closeness centrality, and strategic value under the weights $(0.2,0.3,0.5)$. Node $v_1$ has high degree centrality $0.9$ but low closeness and value ($0.2$, $0.1$), yielding $0.29$; $v_2$ has stronger closeness $0.8$ with moderate degree and value ($0.4$, $0.3$), yielding $0.47$; and $v_3$, despite low degree and closeness ($0.1$, $0.3$), attains the highest score $0.61$ because its strategic value dominates at $1.0$. This illustrates how the weighting combines structural and strategic information when prioritizing candidate attacker objectives.
\end{example}

The resulting strategic ranking highlights the most plausible attacker objectives and subsequently guides graph reduction and AT synthesis.

\subsection{Target-Oriented Graph Reduction}
\label{subsec:graph-reduction}
After a candidate target node $v_{\text{target}}$ or a set of targets $V_{\text{target}}$ is identified through strategic ranking, the defender no longer needs to analyze the entire environment graph. Instead, the analysis can be restricted to nodes and edges that lie on feasible attack paths from the initial observation point $v_{\text{init}}$ to the candidate target $v_{\text{target}}$. This reduction significantly decreases computational overhead and focuses attention on the most relevant portion of the environment.

For demonstration, consider the situation in Table~\ref{tab: table 2}, where an intruder is first observed at $v_2$ at $t_{0}=0$, and the defender evaluates the situation at $t_{\text{now}}=5$ minutes. If $v_3$, with $\deg(v_3)>1$, is ranked as the most probable target, then all nodes located far from $v_3$ and reachable only after approximately $7$ minutes become irrelevant for determining the attacker's current location. Consequently, removing all nodes and edges that do not lie on a feasible path to $v_3$ within the allowed time does not affect the analysis~\cite{ali2025graph}. \emph{Graph reduction} refers to eliminating nodes and edges that cannot contribute to any feasible attack path toward the selected target. Formally, given the elapsed time $\Delta t$, we consider only those paths $p \in \Pi(v_{\text{init}}, v_{\text{target}}, \Delta t)$ whose traversal time satisfies $t_p \leq \Delta t$. All nodes and edges not appearing in any such path are removed from the graph.

The reduced subgraph is formalized as the \emph{target-oriented subgraph}:
\[
G_{\text{sub}} = (V_{\text{sub}}, E_{\text{sub}},\lambda, \Omega_{t}^{c}),
\]
where
\[
V_{\text{sub}} = \bigcup_{p \in \Pi(v_{\text{init}}, v_{\text{target}}, \Delta t)} 
\{\text{nodes in } p\}, \]
\[E_{\text{sub}} = \bigcup_{p \in \Pi(v_{\text{init}}, v_{\text{target}}, \Delta t)} 
\{\text{edges in } p\}.
\]

All labels $\lambda$, weights $\Omega_{t}^{c}$, and observability properties are inherited from the original graph. $V_{\text{sub}}$ contains exactly the nodes that lie on at least one feasible path to the target; that is, $v \in V_{\text{sub}}$ if and only if $d(v_{\text{init}}, v) + d(v, v_{\text{target}}) \leq \Delta t$, where $d(\cdot,\cdot)$ is the shortest traversal time. In the implementation, $E_{\text{sub}}$ is taken as the set of all original edges between retained nodes (the node-induced subgraph). This may retain edges that do not themselves lie on a feasible path to the target, such as back-edges. During attack-tree construction, the enumeration retains only prefixes whose cumulative time does not exceed $\Delta t$ and only complete branches that reach $v_{\text{target}}$. Thus, the generated tree contains no time-infeasible paths.
For every candidate path, construction starts at $v_{\mathrm{init}}$ and maintains the cumulative traversal time of the current prefix. An outgoing edge is excluded from the path when appending it would make the prefix cumulative time exceed $\Delta t$. When a prefix reaches $v_{\mathrm{target}}$, it is retained only if its complete cumulative time is at most $\Delta t$; a prefix that reaches another node without remaining time is pruned. The search enumerates cycle-free paths, because revisiting a node cannot improve a path under positive traversal times. Therefore, edges and branches are removed by the explicit predicate $\sum_{e\in p}t(e)\leq\Delta t$, and every path represented in the generated attack tree satisfies this predicate.

To examine the efficiency gained through reduction, we define the reduction factor over both nodes and edges, following our prior work~\cite{ali2025graph}:
\[
\alpha = \frac{|V_{\text{sub}}| + |E_{\text{sub}}|}{|V| + |E|}.
\]

A value $\alpha < 1$ indicates that the reduced graph contains fewer graph elements than the original graph. The reduction factor therefore serves as a structural measure of the search-space reduction achieved by the proposed method; it should not be interpreted as a direct measurement of end-to-end execution-time improvement. By eliminating nodes and edges that cannot participate in a feasible attack path, the reduction decreases the portion of the environment that must be considered during subsequent attack-tree synthesis. In practice, $\alpha$ often drops substantially, sometimes below $0.2$, especially in large facilities or IoT networks where only a small fraction of nodes lie on feasible attack paths. The experimental evaluation in \Cref{sec:five} demonstrates this reduction across the considered scenarios.

\subsection{Attack Tree Synthesis from Reduced Graphs}
Once the target-oriented subgraph $G_{\mathrm{sub}}$ is obtained, the defender can synthesize an AT that represents the feasible ways in which the attacker may reach the candidate target. Because irrelevant nodes have been removed, the resulting representation is restricted to the temporally feasible portion of the environment graph. 

We adopt the algorithm presented in~\cite{ali2025graph} for AT generation. The canonical AT-generation algorithm enumerates all feasible paths in $G_{\mathrm{sub}}$ from $v_{\mathrm{init}}$ to $v_{\mathrm{target}}$, thereby retaining the feasible attack strategies represented in the reduced subgraph. Each path corresponds to a sequence of actions (edge traversals), and paths that share a common prefix are merged at a common branch point to avoid unnecessary duplication. Common substructures are therefore shared where permitted by the AT construction rather than represented independently for every path. The complexity of the generated AT depends on the number of feasible attack paths in the reduced graph, which may grow exponentially in the worst case. Thus, $\alpha$ provides evidence of search-space reduction rather than a standalone guarantee of computational efficiency.

\section{Implementation and Evaluation}
\label{sec:five}
We implement our framework in Python using the NetworkX\footnote{https://pypi.org/project/networkx/} package and evaluate it across three use cases, covering threat environments ranging from a single building to a multi-building industrial facility and an IoT attack dataset. The code, along with scripts to reproduce the evaluated scenarios, is publicly available online\footnote{https://github.com/Ezecc/Graph2AT4UTA/tree/main}.

\subsection{Experimental Scenarios}

We use three use cases that represent distinct aspects of attack target identification, threat diversity, and deployment on real network data. Each scenario poses challenges in observability and attack-surface complexity.

To distinguish what the method \emph{uses} from what it is \emph{evaluated against}, we state the two explicitly. The algorithm inputs are the environment graph $G$, the observed entry $v_{\text{init}}$, the elapsed time $\Delta t$, the observable set $V_{\text{obs}}$, the ranking weights $w_{(\cdot)}$, and the strategic-value prior $S(v)$. In the building cases the target is inferred: the method ranks candidates from $v_{\text{init}}$ and $\Delta t$, so the true target $v_{\text{target}}$ and its attack path serve only as ground truth for measuring ranking accuracy and coverage. In the IoT case $v_{\text{target}}$ is instead a supplied input (the use case is reduction-only). Consequently, scenario phrases such as ``targeting the data center'' denote the ground-truth objective under evaluation, not information provided to the ranking step.

We first describe the three use cases and then present a baseline comparison against the $k$-hop approach.

\subsubsection{\textbf{Use Case 1: Single Building}}

\begin{figure}
    \centering
    \includegraphics[width=1\linewidth]{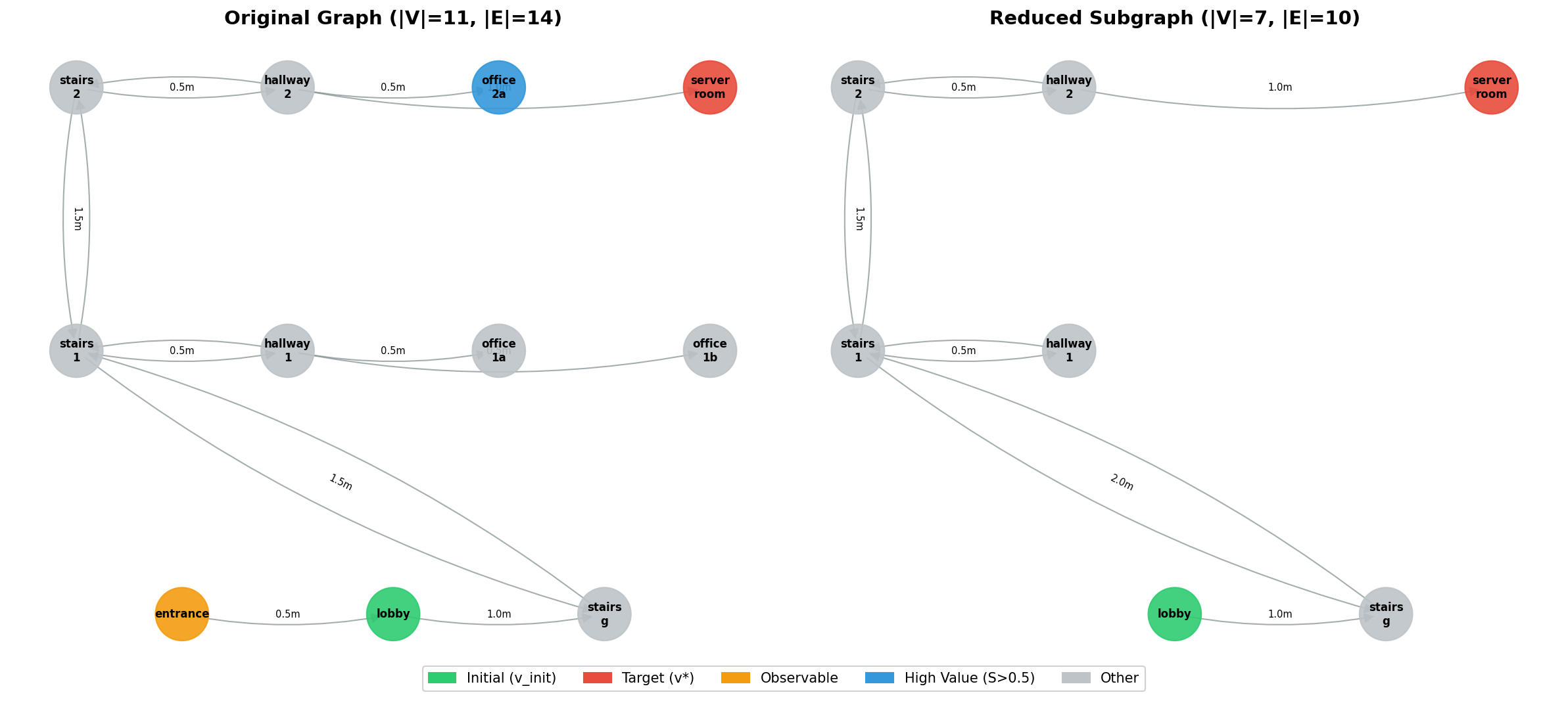}
    \caption{Single-building graph and its reduced version after applying temporal constraints from an observation point}
    \label{fig:Single-Building-gph-red}
\end{figure}

The first scenario models a single three-floor office building as a graph with 11 nodes and 14 directed edges representing locations and transitions, respectively. The building consists of a ground floor with an entrance and a lobby. We assume that an intruder can be observed at some of these locations, such as the lobby, but subsequent movement is unobservable. The left side of Fig.~\ref{fig:Single-Building-gph-red} illustrates the single-building graph, whose nodes represent the first floor with two offices and a hallway and the second floor with an executive office and a server room. Each floor is connected via staircases, and traversal times and generalized costs are assigned to the edges. The costs are dimensionless priors in $[0.1, 0.5]$ and are larger for transitions into sensitive spaces: $0.1$ for entrance and hallway moves, $0.2$--$0.3$ for stairwells, $0.2$ into the first-floor offices, $0.3$ into the executive office, and $0.5$ into the server room.

Strategic values are assigned to reflect the criticality of the nodes, for example the server room (\texttt{server\_room}) receiving the highest value $1.0$, the executive office (\texttt{office\_2a}) $0.9$, and other locations receiving lower values proportional to their importance. For the primary test case, we set the initial observation point to the lobby with an elapsed time of $8.0$ minutes. We assume that an attacker is detected entering through the main entrance and moving into the lobby. From this observable position, the defender must determine which high-value target the attacker is likely to pursue. The temporal reachability analysis computes $\mathcal{R}^+(\texttt{lobby}, 8.0)$, the set of nodes the intruder could reach from the lobby within $8.0$ minutes. The ranking identifies $\texttt{server\_room}$ as the top candidate target due to its combination of the highest strategic value and reachability within the time constraint.

The graph reduction procedure constructs the target-oriented subgraph $G_{\text{sub}}$ by extracting all feasible paths from the $\texttt{lobby}$ to the $\texttt{server\_room}$ that satisfy $t_p \leq 8.0$ minutes. The reduced subgraph is shown on the right-hand side of Fig.~\ref{fig:Single-Building-gph-red}. The reduction yields a subgraph with 7 nodes and 10 edges, achieving a reduction factor of $\alpha = 0.68$. This represents a 36\% reduction in nodes while preserving all paths necessary to reach the identified target.

\begin{figure}
    \centering
    \includegraphics[width=1\linewidth]{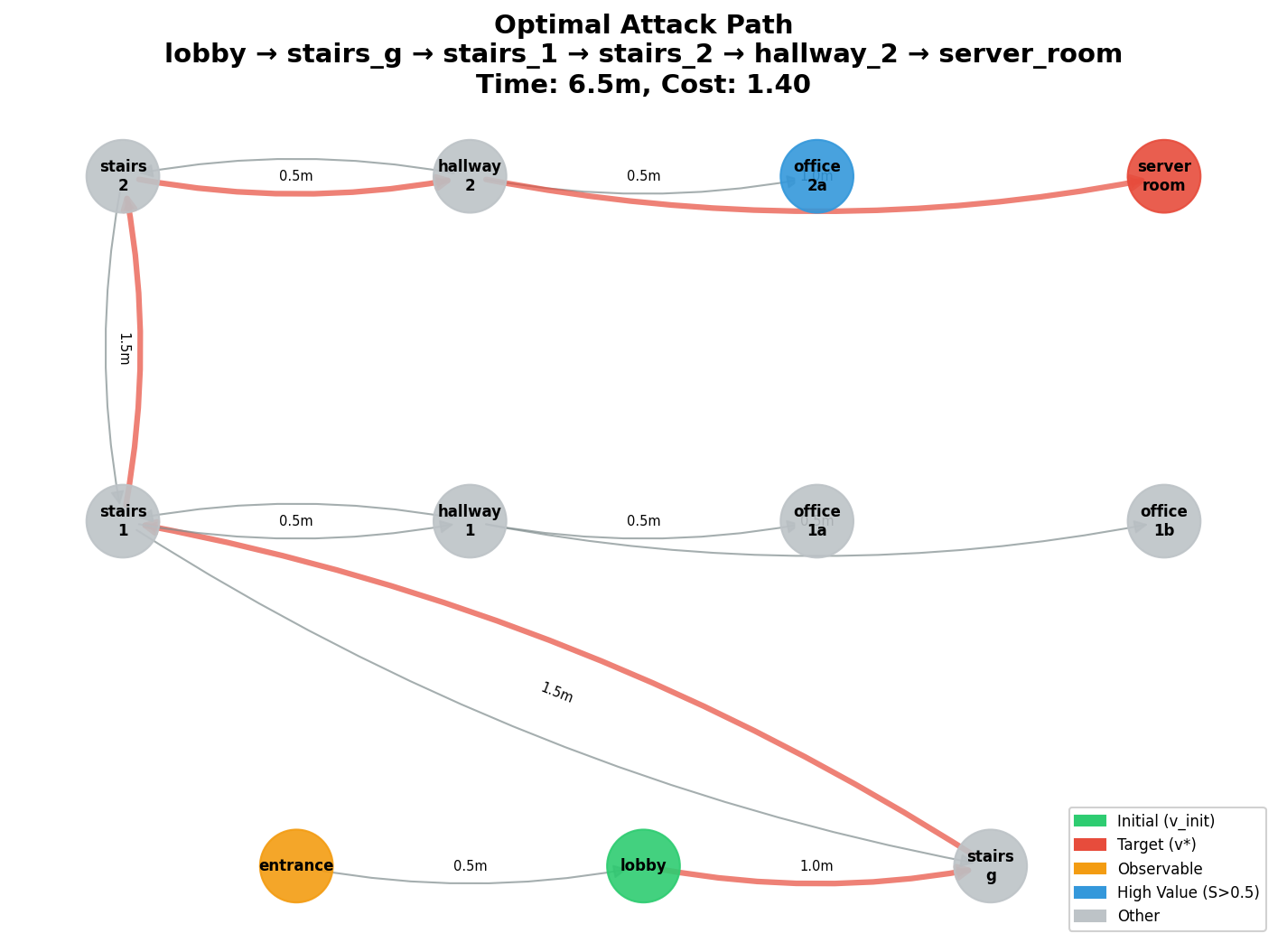}
    \caption{Single-building layout and optimal-path preservation}
    \label{fig:Single-Building-optml-path}
\end{figure}

\begin{figure}
    \centering
    \includegraphics[width=1\linewidth]{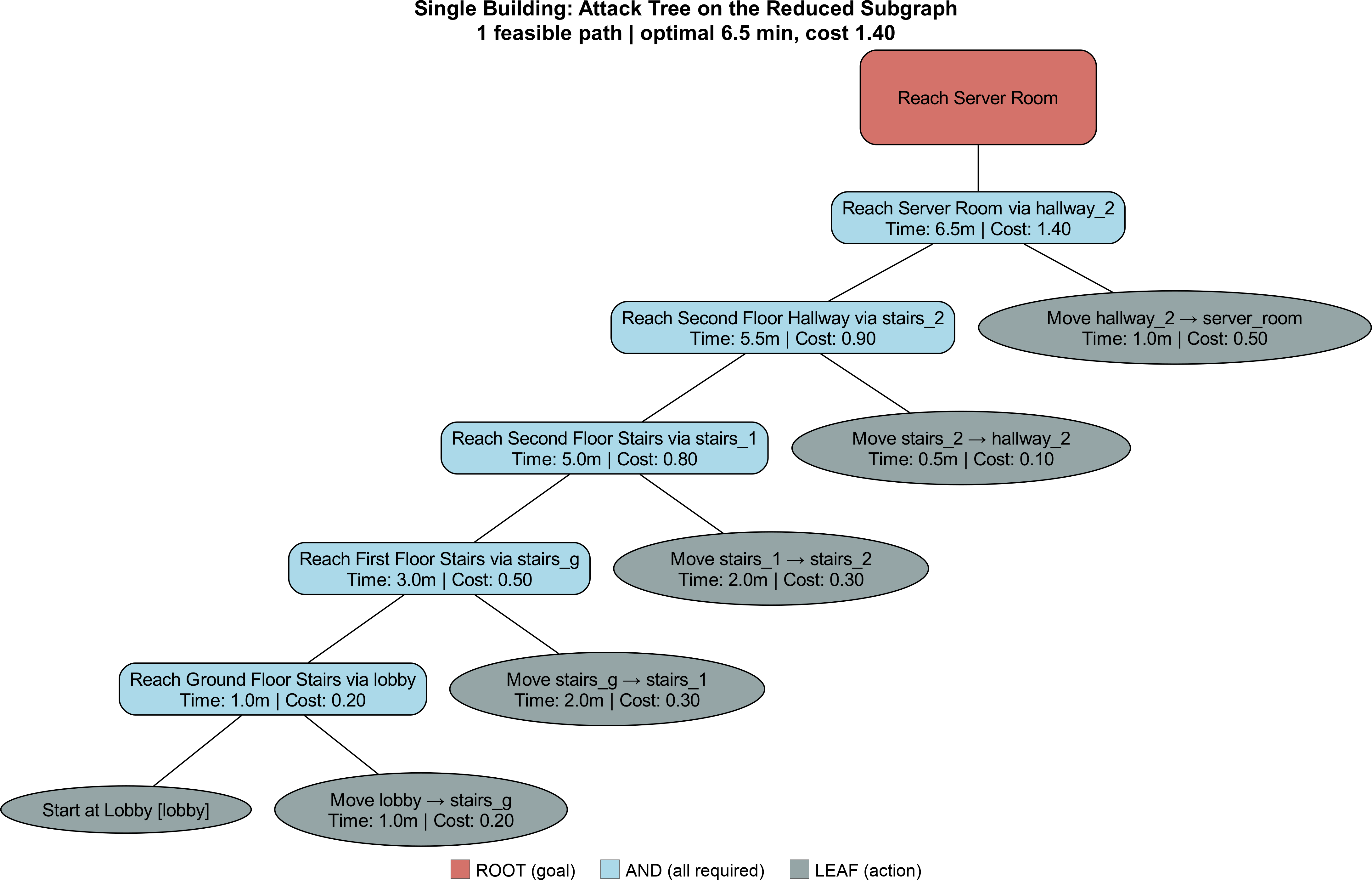}
    \caption{An attack tree containing a sequential-AND structure whose nodes must be traversed in order to reach the target.}
    \label{fig:Single-Building-AT}
\end{figure}

Only one cycle-free path leads from the lobby to the identified target, so it is also the optimal path, as shown in Fig.~\ref{fig:Single-Building-optml-path}. The path includes the sequence \{\texttt{lobby}, \texttt{stairs\_g}, \texttt{stairs\_1},
\texttt{stairs\_2}, \texttt{hallway\_2}, \texttt{server\_room}\}
with a total traversal time of 6.5 minutes and a cumulative cost of $1.40$, the sum of its five edge costs ($0.2 + 0.3 + 0.3 + 0.1 + 0.5$). Because the cost constraint is not enforced, this value is an attribute of the resulting AT and does not affect feasibility. Based on this reduction and target identification, an AT (shown in
Fig.~\ref{fig:Single-Building-AT}) is generated that shows how the intruder could reach the target. This includes a sequence of steps that lead to the target, represented
as a sequential-AND structure in which the nodes must be traversed in order to reach the target.

\subsubsection{\textbf{Use Case 2: Multi-building Industrial facility}}

\begin{figure*}[!ht]
    \centering
    \includegraphics[width=1\linewidth]{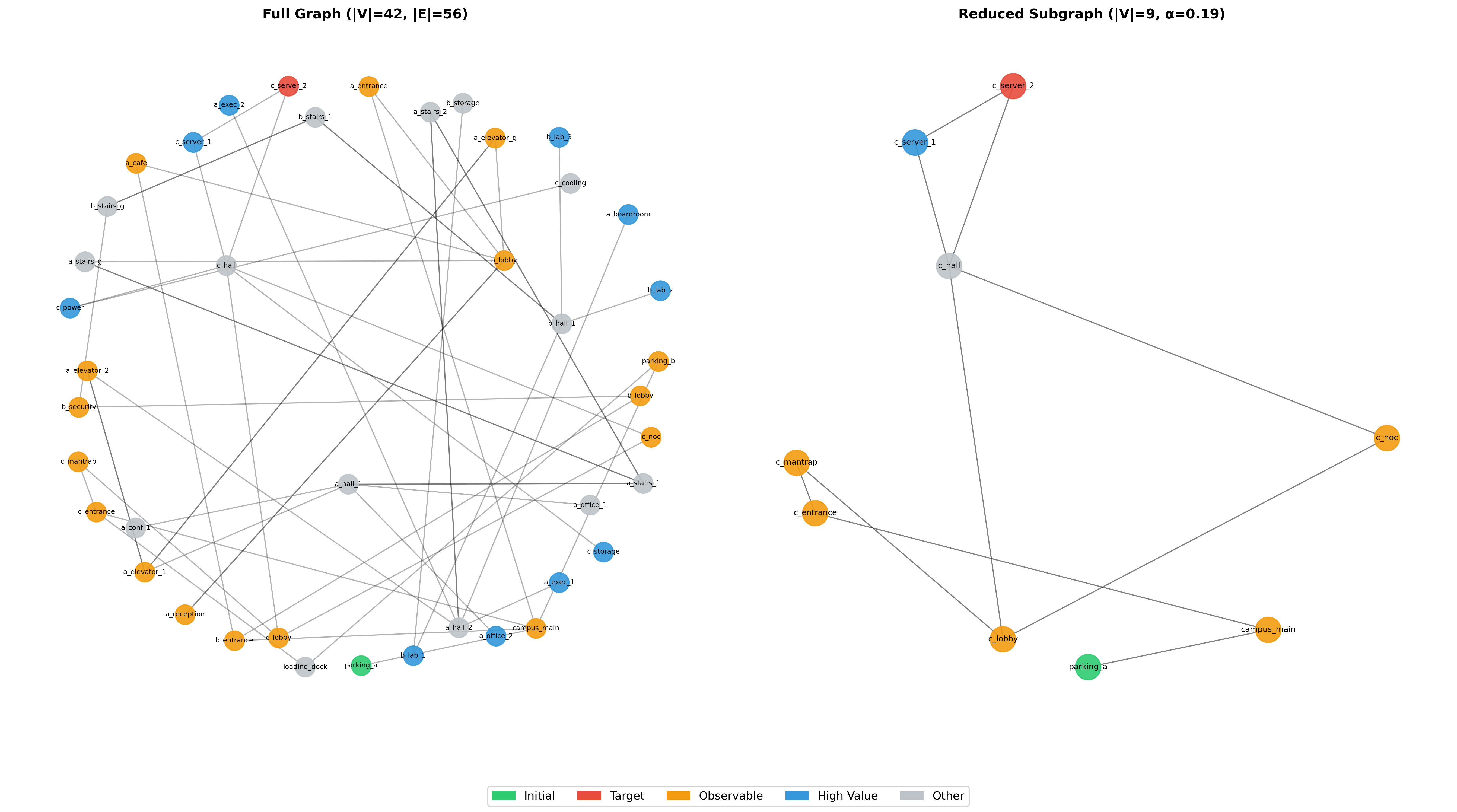}
    \caption{Multi-building industrial facility layout}
    \label{fig:multiple-Building}
\end{figure*}

The second scenario scales up to a multi-building industrial facility with 42 nodes and 56 directed edges, representing an environment four times larger than the single-building case. Fig.~\ref{fig:multiple-Building} shows the full graph on the left and the reduced graph on the right. The building facility consists of three interconnected structures, A, B, and C.

Building A houses administrative functions across 18 nodes including office spaces, an executive suite, reception areas, and vertical transport via elevators. Building B contains research laboratories spanning 10 nodes with chemistry, biology, and physics labs plus equipment storage. Building C serves as the data center with 10 nodes including primary and secondary server rooms, network operations center, power infrastructure, and backup storage. The buildings are connected through outdoor pathways with 4 nodes representing parking areas, campus access points, and service entrances. As in the single building, each edge carries a traversal time and a dimensionless cost prior. The costs range from 0.1 for outdoor walkways and most corridor moves to 0.5--0.6, the highest values, which occur only on transitions into secured spaces: the data-center mantrap, the server rooms, and the executive office.

In practice, observable locations typically correspond to areas already covered by security instrumentation such as cameras, badge readers, or security personnel; for IoT deployments, observability is instead determined by the availability of device telemetry, logs, or monitoring services. The partition is therefore a deployment fact rather than a modeling choice: a defender reads it off the existing monitoring inventory. In this case study, 17 locations are designated observable: all building entrances, lobbies, and selected high-traffic areas covered by cameras or security personnel. An attacker entering through any of them can be detected. More generally, the observability function $\mathcal{O}(v,t)$ can accommodate time-varying coverage (e.g.\ sensor outages or maintenance), but we assume a static deployment for clarity. High-value targets are distributed throughout the facility: the primary data-center server room in Building~C is rated $\texttt{c\_server\_2} = 0.95$, the executive office $\texttt{a\_exec\_1} = 0.90$, and critical research laboratories $0.70$--$0.80$.

\begin{figure*}[!ht]
    \centering
    \includegraphics[width=\linewidth]{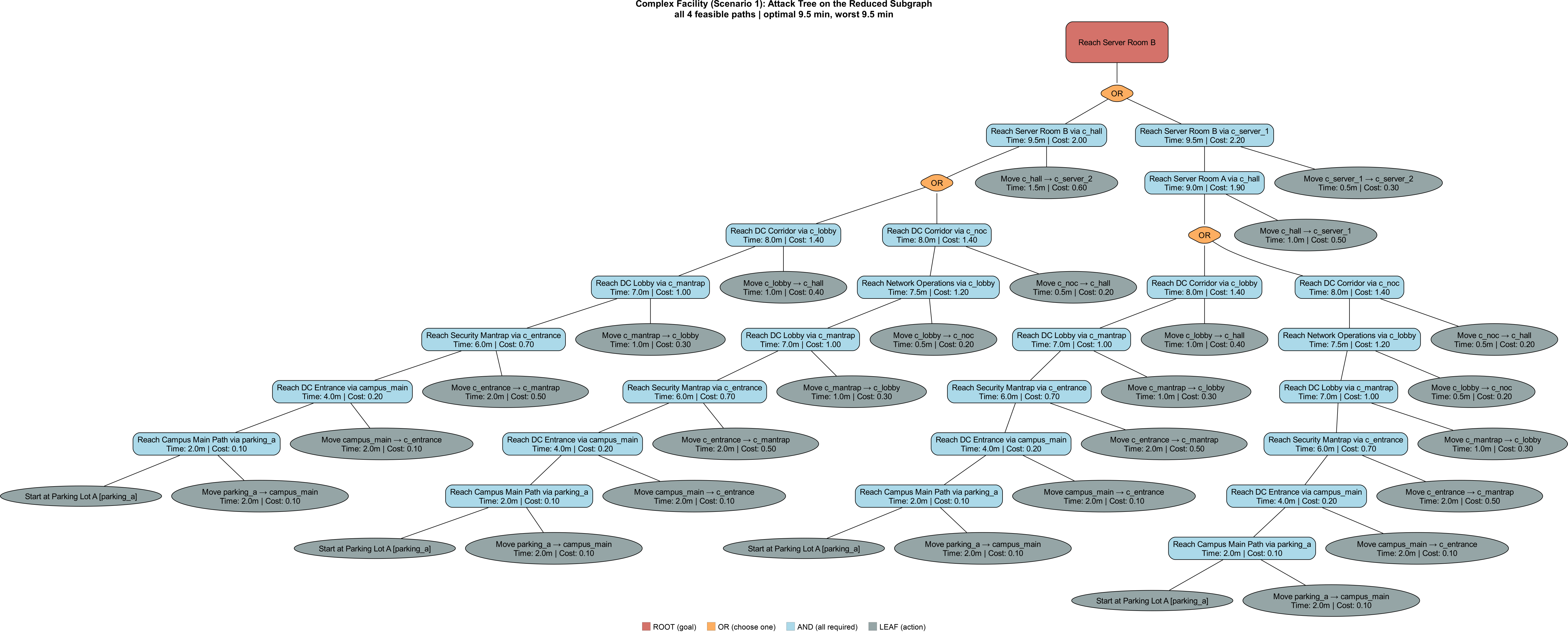}
    \caption{Attack tree for complex facility Scenario~1, synthesized from the reduced subgraph. It enumerates all four feasible routes to the data center as OR-alternatives (all four take 9.5 minutes and differ only in cost), rather than only the optimal path.}
    \label{fig:multiple-Building-tree}
\end{figure*}

We evaluate five attack scenarios spanning different threat vectors and operational constraints. Scenario~1 models an external attacker first observed in the parking area $15.0$ minutes earlier, targeting the primary data center; Scenario~2 a similar external threat toward the executive offices, observed $12.0$ minutes earlier; Scenario~3 an insider from the Building~A administrative lobby heading for the Building~B biology lab, observed $10.0$ minutes earlier; Scenario~4 campus-perimeter access through the main entrance toward the secondary server room, observed $10.0$ minutes earlier; and Scenario~5 exploitation of the service entrance toward the power infrastructure, observed $8.0$ minutes earlier. In every scenario the target is unknown at the time of observation. As each scenario yields its own reduced subgraph and AT, we present the reduction and AT for the first scenario and provide the rest online\footnote{https://github.com/Ezecc/Graph2AT4UTA/tree/main}.

For the first attack scenario, the temporal reachability analysis from the parking area computes $\mathcal{R}^+(\texttt{parking\_a}, 15.0)$ which includes nodes on paths traversing outdoor walkways, building entrances, and vertical circulation to reach upper floors. Using the default weights ($w_D=0.2$, $w_C=0.3$, $w_S=0.5$), the strategic ranking identifies $\texttt{c\_server\_2}$ as the top-ranked candidate ($R(\texttt{c\_server\_2}) = 0.50$, ahead of $\texttt{a\_exec\_1}$ at $0.49$), reflecting its combination of the highest strategic value in the facility and reachability within the window.

The graph reduction for this scenario yields $\alpha = 0.19$, retaining only 9 of the original 42 nodes (a 79\% reduction in nodes and 81\% in graph elements) while maintaining all feasible paths to the target (Fig.~\ref{fig:multiple-Building}, right). The $15.0$-minute constraint eliminates the many peripheral locations that cannot be reached within the operational window. The minimum-time path is
\{\texttt{parking\_a}, \texttt{campus\_main}, \texttt{c\_entrance},
\texttt{c\_mantrap}, \texttt{c\_lobby}, \texttt{c\_hall},
\texttt{c\_server\_2}\}
with a traversal time of 9.5 minutes, passing through the data center's security mantrap and internal corridor.

The ATs generated for the multi-building industrial facility scenarios maintain the same structural properties as the single building case. For the first attack scenario, the AT is presented in Fig.~\ref{fig:multiple-Building-tree}. The tree enumerates all four feasible routes to the identified target; all take 9.5 minutes, and the two routes via $c\_server\_1$ cost more (2.20 vs 2.00).

\subsubsection{\textbf{Use Case 3: IoT attack dataset}}

\begin{figure*}
    \centering
    \includegraphics[width=1\linewidth]{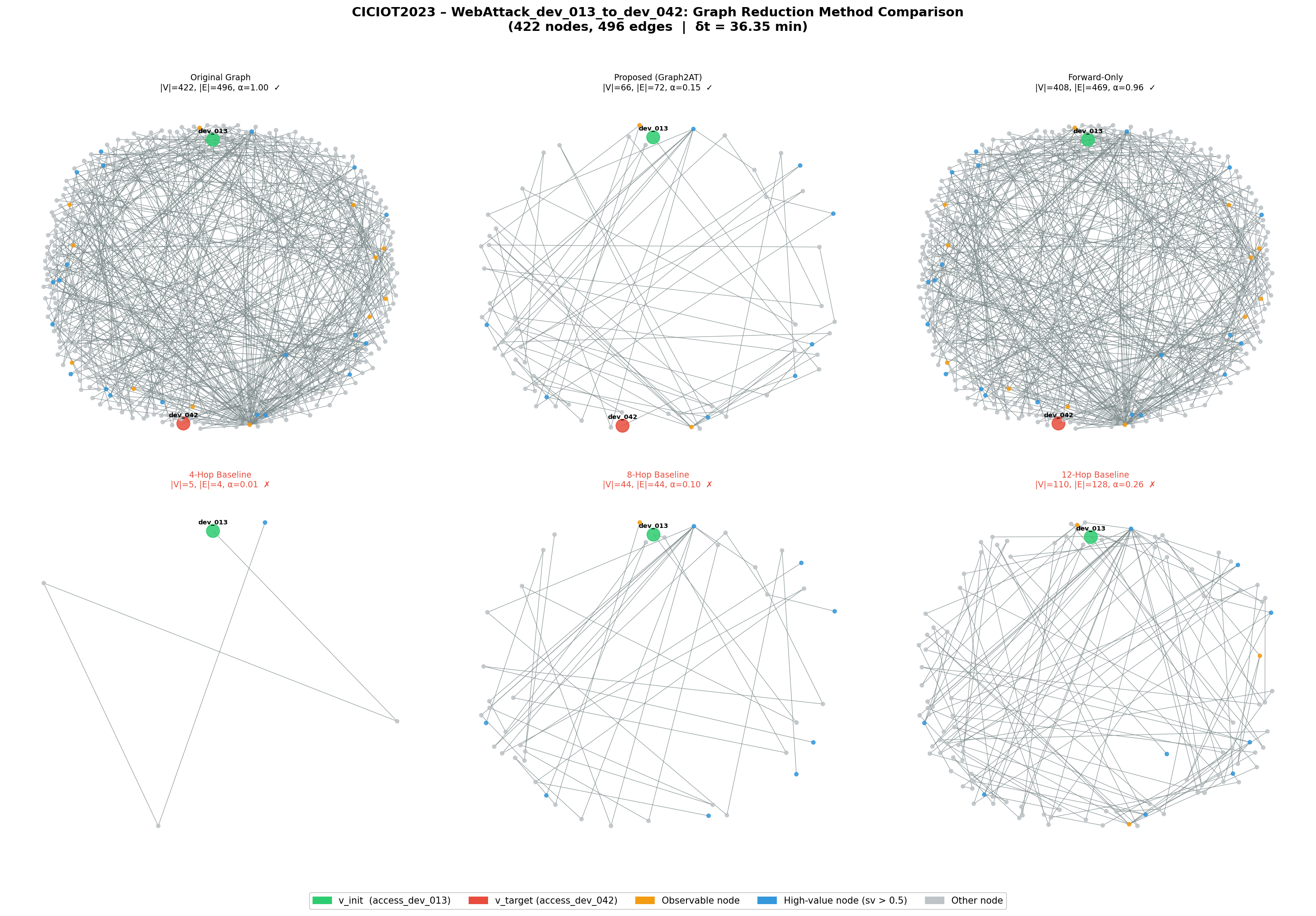}
    \caption{Graph reduction for the CICIoT2023 \texttt{WebAttack} scenario ($\texttt{dev\_013}\rightarrow\texttt{dev\_042}$, $T_{path}=36.35$ min) on the real device-communication topology. Top row: the original state-based graph ($|V|=422$), the proposed reduction ($\alpha=0.15$), and the forward-only reachable set ($\alpha=0.96$). Bottom row: the $k$-hop baselines for $k=4,8,12$. The initial node is shown in green and the supplied target in red; only the proposed reduction retains the target while remaining compact.}
    \label{fig:IoT-use-case-Graph}
\end{figure*}

The third use case evaluates the proposed methodology on the CICIoT2023 dataset~\cite{neto2023ciciot2023}. This benchmark models 33 attacks in an IoT environment composed of 105 devices. The attacks fall into seven categories: DDoS, DoS, Recon, Web-based, Brute Force, Spoofing, and Mirai. The dataset also contains non-attack (benign) samples. All attacks are executed by malicious IoT devices targeting other IoT devices. We build the environment graph from the dataset's \emph{real} device-communication topology, which we recover from the internal device-to-device traffic of the benign packet capture: every pair of hosts observed communicating becomes a directed link, yielding $50$ active devices and $124$ directed links. Device vendors are resolved from the IEEE OUI registry (all $50$ MAC prefixes resolve), and device types are inferred from the vendor. Expanding each device into its access state and each link into its attack-step transitions (vulnerability discovery, exploitation, lateral movement) yields a state-based graph of $422$ nodes and $496$ directed edges. This graph is shown in Fig.~\ref{fig:IoT-use-case-Graph}.
More specifically, the expansion is constructed in two layers. The 50 recovered device hosts provide the base device states, and each of the 124 directed communication links is expanded into three attack-step transitions---vulnerability discovery, exploitation, and lateral movement---with corresponding intermediate states and directed edges. Thus, the state graph contains $50 + 3(124)=422$ nodes and $124 + 3(124)=496$ directed edges. The three attack-step stages are therefore modeled state-transition abstractions derived from the communication topology; they are not reconstructed from individual CICIoT2023 attacker--target traces.

\begin{figure*}
    \centering
    \includegraphics[width=1\linewidth]{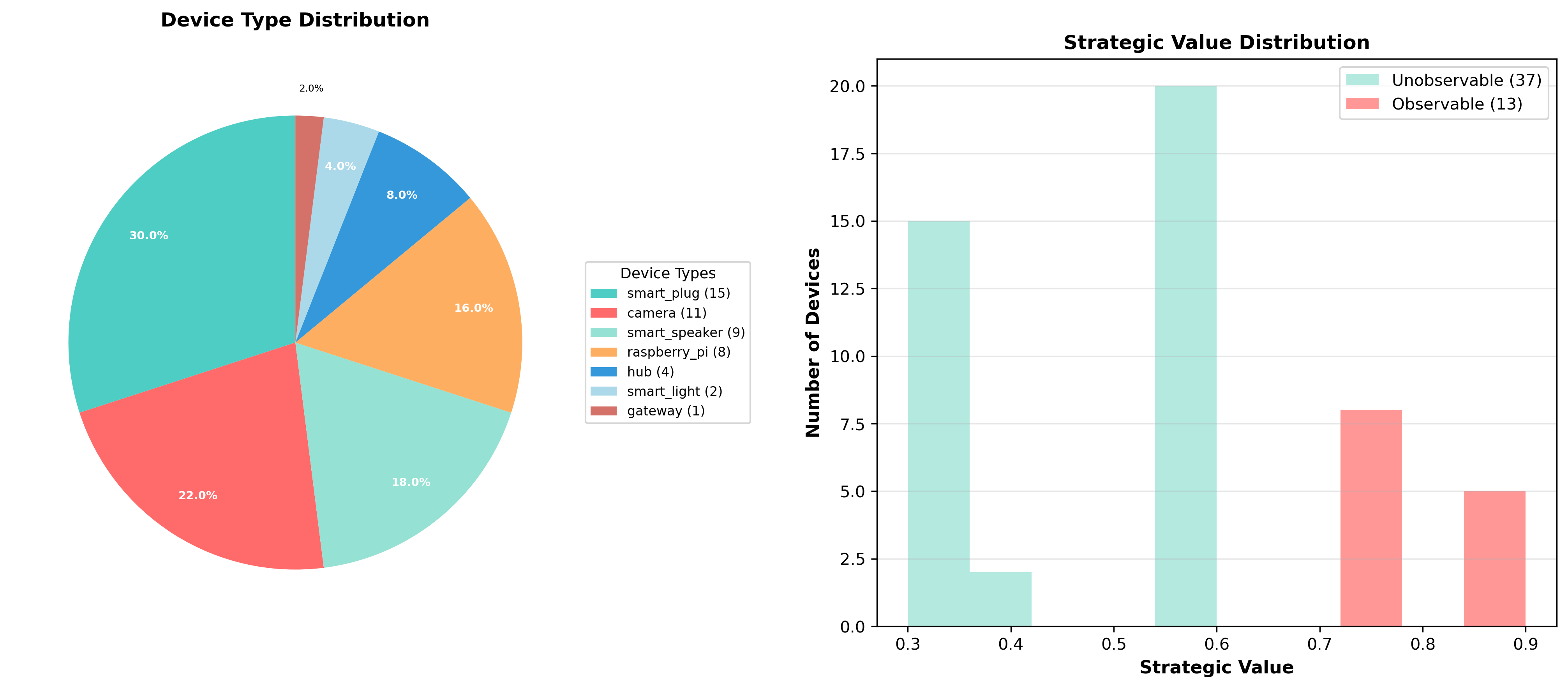}
    \caption{Device types and strategic value distributions.}
    \label{fig:iot_devices_and_strategic_values}
\end{figure*}

We assign strategic values based on device type criticality: the \texttt{gateway} is assigned $0.90$, \texttt{hubs} $0.85$, \texttt{Raspberry Pi servers} $0.75$, \texttt{cameras} $0.60$, \texttt{smart speakers} $0.55$, \texttt{smart lights} $0.40$, and \texttt{smart plugs} $0.30$. Observable nodes are limited to the gateway, hub devices, and Raspberry Pi servers ($|V_{\text{obs}}| = 13$), reflecting realistic IoT deployments where low-end devices lack monitoring capabilities as shown in Fig.~\ref{fig:iot_devices_and_strategic_values}. The strategic values, observability assignment, edge traversal times, and edge cost are defender-side modeling priors rather than measured dataset quantities, and we disclose them as such. We model edge traversal time rather than measuring it. In the building use cases (physical security), traversal time is generally proportional to path distance; in the IoT use case (cyber), it instead reflects the effort of compromising each hop. Concretely, we model each hop's time to increase with the criticality of the device being compromised (better-defended, higher-value devices take longer to breach), plus an attack-type-dependent dwell-time buffer $b$. The dataset cannot supply this timing: its attack records are aggregated flow windows with packet identity stripped, yielding neither attacker$\rightarrow$target pairs nor per-step durations. Moreover, packet-transmission time is not the attacker's dwell time, since a flood spans seconds of packets but minutes of reconnaissance, exploitation, and pivoting. Attackers may also interleave benign steps to evade detection~\cite{sen2022holistic}.
Therefore, we assign $t_i$ and derive $T_{path}$ synthetically for each scenario as follows.
\begin{equation*}
    T_{path} = \sum_{i=1}^{H} t_i + b
\end{equation*}
where $H$ is the number of hops on the ground-truth shortest path, $t_i$ is the modeled edge traversal time of hop $i$, and $b$ is an attack-type-dependent dwell buffer. We set $b$ according to the differing dwell and preparation characteristics of each attack class: fast flooding attacks are given a short buffer (DoS and DDoS, $3$ min), reconnaissance and spoofing intermediate values ($4$--$5$ min), web-based compromise $6$ min, and slow credential or botnet campaigns the longest (BruteForce $10$ min, Mirai $12$ min). A minutes-scale per-hop budget is consistent with empirical reports that compromises are typically measured in minutes~\cite{verizon2017dbir} and that initial-access-to-lateral-movement (``breakout'') times average tens of minutes~\cite{crowdstrike2026gtr}. The specific per-attack-class values remain a disclosed modeling choice; only their relative ordering (floods fast, credential/botnet campaigns slow) is intended to be meaningful. This yields a realistic upper bound that the optimal path must fit within while remaining tight enough to make the reduction non-trivial. The use case thus demonstrates reduction on a \emph{real} topology under \emph{modeled} timing. The conclusions do not hinge on the exact timing. Perturbing every edge time by up to $\pm 50\%$ (five random trials per level) leaves target preservation at $100\%$ across all ten scenarios and the average reduction factor unchanged at $\approx 0.056$. Coverage is in fact invariant by construction, since the optimal path defines the budget and always remains feasible. In each perturbation trial, the time budget is recalculated from the perturbed edge times: the perturbed shortest-path time $H$ is recomputed for the supplied target, and $T_{\mathrm{path}}$ is set to the resulting sum plus the unchanged attack-class dwell buffer $b$. The original, unperturbed $T_{\mathrm{path}}$ is therefore not held fixed. Consequently, this experiment establishes preservation when the temporal constraint is reconstructed around each perturbed optimum; it does not by itself establish robustness under a fixed, externally specified time budget. Edge costs are modeled in the same way as traversal times: a device hop into device $v$ receives the dimensionless cost $c = 0.5\,S(v) + \epsilon$, with $\epsilon \sim \mathcal{U}(-0.05, 0.10)$ (range $0.10$--$0.55$), split across the hop's attack-step transitions. Similar to the building cases, costs annotate the AT steps but do not exclude paths.

We constructed 10 IoT attack scenarios on this real topology, each defined by an initial observation point and a supplied target host. All ten were evaluated. Herein, we present the \texttt{WebAttack} scenario (a web-based compromise propagating from a smart plug toward a Raspberry Pi server) as a representative example, because it is the only scenario whose ground-truth path spans four device hops, which makes the reduction non-trivial.

\paragraph{WebAttack scenario description}

This models a web attack originating from a Belkin smart plug (device $\texttt{dev\_013}$) and targeting a Raspberry Pi server (device $\texttt{dev\_042}$). On the real topology the shortest path traverses four device hops,
\[
\resizebox{\columnwidth}{!}{$\texttt{dev\_013} \rightarrow \texttt{dev\_028} \rightarrow \texttt{dev\_011} \rightarrow \texttt{dev\_015} \rightarrow \texttt{dev\_042},$}
\]
that is, 17 states and 16 state transitions through two smart speakers and the gateway. The modeled device-level traversal time is $30.35$ minutes; adding the $6$-minute dwell buffer for web-based attacks yields a temporal constraint $T_{path} = 36.35$ minutes. As in previous scenarios, the attacker is initially observed at the compromised source device, after which subsequent movements are considered unobservable unless the underlying device belongs to the observable set (gateway, hubs, and Raspberry Pi servers).

\paragraph{Supplied target and scope of inference}
In the building use cases the target is \emph{inferred}: the strategic ranking $R(v)$ identifies a uniquely dominant high-value asset (e.g.\ the server room). The IoT use case differs: we fix the target from the scenario's known victim host and apply the reduction only. This scoping is deliberate and structural. The recovered topology is gateway-centric: roughly $90\%$ of devices route through the gateway, so it dominates every centrality measure. $R(v)$ therefore returns it for every observation point, a constant rather than a per-scenario predictor. The dataset compounds this: it documents victims per \emph{attack type} but no attacker source or strategic target, leaving no independent ground truth to infer against. Notably, the \texttt{WebAttack} target ($\texttt{dev\_042}$, $S=0.75$) is not the highest-value node (the gateway, $0.90$), so a value-only heuristic would mis-rank it, which is precisely why the use case relies on the supplied victim. The IoT use case therefore demonstrates \emph{graph reduction with a supplied target}; we return to the prospect of inference on such topologies in Section~\ref{sec:conclusion}.

\begin{figure*}
    \centering
    \includegraphics[width=1\linewidth]{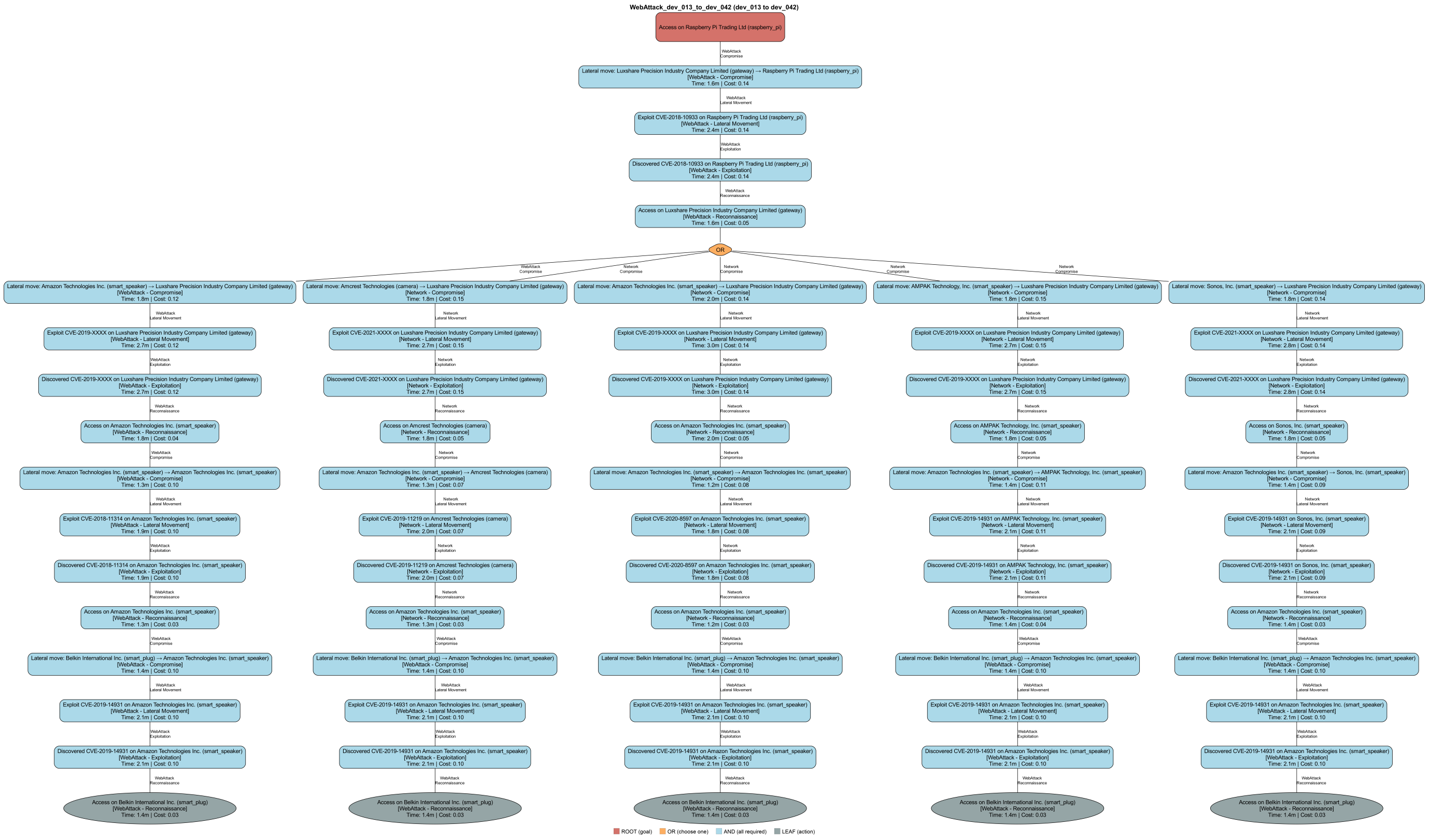}
    \caption{Attack tree for the CICIoT2023 \texttt{WebAttack} scenario ($\texttt{dev\_013}\rightarrow\texttt{dev\_042}$), constructed from the proposed reduced subgraph.}
    \label{fig:IoT-use-case-AT}
\end{figure*}

Fig.~\ref{fig:IoT-use-case-Graph} shows the reduction for this scenario; we construct the AT from the proposed target-oriented subgraph (top-center), shown in Fig.~\ref{fig:IoT-use-case-AT}.

\subsection{Baseline Comparison with $k$-hop}

To evaluate the effectiveness of temporal reachability reduction, we compare the proposed method against a classical $k$-hop neighborhood baseline \cite{nikolentzos2020k}. The $k$-hop method constructs a subgraph via topology-only breadth-first search (BFS), including all nodes within $k$ edges of the initial observation node, independent of time constraints and target reachability.

For a given initial node $v_{\text{init}}$ and parameter $k \in \mathbb{N}$, the $k$-hop neighborhood $N_k(v_{\text{init}})$ is defined as 
\begin{equation*}
    N_k(v_{\text{init}}) = \{v \in V \mid d_{\text{hop}}(v_{\text{init}}, v) \leq k\}, 
\end{equation*} 

where $d_{\text{hop}}(u, v)$ denotes the minimum number of edges in any path from $u$ to $v$. The induced subgraph $G_k$ contains all nodes in $N_k(v_{\text{init}})$ and all edges with both endpoints in this set. The reduction factor is $\alpha_k = (|N_k| + |E_k|) / (|V| + |E|)$, where $E_k$ is the edge set of $G_k$. The $k$-hop approach has two limitations: it ignores time constraints (admitting nodes that are temporally unreachable within $T_{path}$) and the target location (retaining nodes on no path toward the target), and both worsen as graph size and path diversity grow.

For the \texttt{WebAttack} scenario, whose ground-truth path spans four device hops (16 state transitions), the proposed method attains $\alpha = 0.150$ with full coverage, whereas $k$-hop must reach $k=16$ before it includes the target (at $k=4,8,12$ it retains $\alpha = 0.010, 0.096, 0.259$ yet still misses it). The gain comes from intersecting the forward and backward time-bounded reachable sets, which eliminates nodes within $k$ hops that cannot reach the target within $T_{path}$.

Across all ten IoT scenarios the proposed method averages $\alpha = 0.056$ at $100\%$ coverage. No small $k$-hop setting matches this coverage: $k=4$ covers $0/10$ ($\alpha = 0.012$), $k=8$ covers $8/10$ ($0.372$), and $k=12$ covers $9/10$ ($0.768$); only $k=16$ reaches $10/10$, at an average $\alpha = 0.945$ that retains almost the entire graph. Thus even when tuned to match coverage, $k$-hop retains an order of magnitude more of the graph, because it cannot exclude the temporally infeasible paths that the time-and-target-aware reduction removes. Unlike $k$-hop, the proposed approach does not require manual tuning of $k$, adapts automatically to time constraints, and preserves optimal-path semantics by construction.

Because $k$-hop is a pure-topology method, we additionally conduct two controlled ablations comparing the proposed method against the \emph{forward-only} and \emph{all-paths} variants. The \emph{forward-only} variant keeps every node time-reachable from $v_{\text{init}}$ within $T_{\text{path}}$ (time-aware and target-agnostic); the \emph{all-paths} variant keeps every node on some path from $v_{\text{init}}$ to the target irrespective of time (target-aware and time-agnostic). Both achieve full coverage but provide less reduction: for IoT they average $\alpha = 0.81$ and $0.95$ respectively (and $\alpha = 0.77$ for forward-only on the building cases), versus $\alpha = 0.056$ for the proposed method, which alone exploits time and target jointly. We note that these variants are also the approach used to instantiate the reduction families employed by recent attack-graph optimization work: importance/path-based pruning~\cite{bai2025network} and large-scale node pruning for operational-technology graphs~\cite{viticchie2025advanced}. Neither incorporates temporal feasibility or unobservable-target inference.

\subsection{Target-Ranking Baselines}
\label{sec:ranking-baselines}
Separately from reduction, we evaluate the strategic ranking $R(v)$ that selects the candidate target against four ranking baselines: degree centrality, PageRank, value-only, and random selection. On the single building, where one asset (\texttt{server\_room}, $S=1.0$) dominates, $R(v)$ identifies the correct target and remains stable across all weight configurations with $w_S$ $\ge$ 0.2 (Section~\ref{sec:sensitivity}). On the complex facility, which contains several assets rated $\geq 0.8$, we report top-1, top-3, and mean reciprocal rank (MRR) across the five scenarios, since a single designated target is no longer the only plausible objective; in this regime the ranking is best read as a defender \emph{shortlist}. These ranking metrics are distinct from the target-preservation metric reported in Section~\ref{sec:overall-performance}. Top-1, Top-3, and MRR measure the position of the designated target in the candidate ranking, whereas target preservation measures whether the target supplied to the graph-reduction stage is retained in the resulting target-oriented subgraph. Thus, target preservation evaluates the reduction step, not target-ranking accuracy.

\begin{table}[t]
\centering
\caption{Target-ranking comparison on the complex facility (five scenarios).}
\label{tab:ranking-baselines}
\begin{tabular}{l c c c}
\hline
\textbf{Method} & \textbf{Top-1} & \textbf{Top-3} & \textbf{MRR} \\
\hline
Proposed $R(v)$ & 1/5 & \textbf{4/5} & 0.48 \\
Value-only & 1/5 & \textbf{4/5} & 0.48 \\
Degree centrality & 0/5 & 1/5 & 0.25 \\
PageRank & 1/5 & 1/5 & 0.34 \\
Random (expected) & 0.10 & 0.30 & 0.29 \\
\hline
\end{tabular}

\smallskip
{\footnotesize\raggedright \textit{Note:} Top-1/Top-3 for the deterministic methods are counts over the five scenarios; \textit{Random} is the analytic expectation of a uniformly random ranking (a single random draw is seed-dependent), reported as probabilities.\par}
\end{table}

As Table~\ref{tab:ranking-baselines} shows, $R(v)$ places the designated target in the top three in four of five scenarios, matching the strongest single-signal baseline (value-only) and outperforming centrality, PageRank, and random selection. We do not claim top-1 dominance over value-only on flat multi-target environments: when values are well calibrated, value-only is a strong special case of $R(v)$. The advantage of $R(v)$ is robustness: it combines value with reachability-aware centrality to give correct identification when a dominant target exists (single building) and a reliable shortlist when several high-value assets compete (complex facility).
The centrality terms do not provide an improvement in the five reported top-1/top-3 outcomes; those outcomes are tied to value-only because strategic value is the dominant signal in this evaluation. Their additional benefit is to provide reachability-aware discrimination when strategic values are tied or imperfectly calibrated: degree centrality favors structurally influential locations, whereas time-weighted closeness favors locations reachable quickly from observed positions. Consequently, $R(v)$ is a more general operating rule than value-only, combining defender-supplied importance with temporal and structural information. The ablation and sensitivity results support this as a robustness feature rather than as a demonstrated independent gain in Top-1 accuracy.

 \section{Results and Analysis}
\label{sec:six}

We present results across the building-security and IoT network attack scenarios. The achieved reduction depends on graph structure and target position, as detailed below, while target preservation remains complete in every case.

\subsection{Overall Performance Metrics}\label{sec:overall-performance}

Table~\ref{tab:results_summary} summarizes key metrics across all three use cases. The reduction factor $\alpha$ ranges from 0.14 to 0.68 in the building cases and averages 0.06 on the IoT topology, depending on graph structure and the target's position: the reduction is strongest when the target lies behind a narrow set of feasible paths, so that most of the graph is eliminated. The complex facility scenarios reduce more than the single building because the larger graph contains more peripheral nodes that can be safely removed while preserving paths to the target.

\begin{table}[!t] \centering
\caption{Performance comparison across use cases}
\label{tab:results_summary}
\begin{small} \resizebox{\columnwidth}{!}{\begin{tabular}{l c c c c c}
    \hline
    \textbf{Use Case} & \textbf{Nodes} & \boldmath$\alpha$ & \textbf{Preservation} & \textbf{Optimal Time} & \textbf{Best $k$-hop \boldmath$\alpha$} \\
    \hline
    Single Building & 11 & 0.68 & 100\% & 6.5 min & 0.92 \\
    Complex (S1) & 42 & 0.19 & 100\% & 9.5 min & 0.65 \\
    Complex (S2) & 42 & 0.37 & 100\% & 6.5 min & 0.93 \\
    Complex (S3) & 42 & 0.22 & 100\% & 9.0 min & 0.63 \\
    Complex (S4) & 42 & 0.14 & 100\% & 7.0 min & 0.63 \\
    Complex (S5) & 42 & 0.14 & 100\% & 6.0 min & 0.21 \\
    IoT (WebAttack) & 422 & 0.15 & 100\% & 30.4 min & 0.71$^{*}$ \\
    IoT (avg) & 422 & 0.06 & 100\% & 19.7 min & 0.95$^{*}$ \\
    \hline
    \end{tabular}
} \end{small}
\end{table}

Target preservation remains at 100\% across all scenarios, meaning that the target supplied to the reduction stage is retained in every resulting target-oriented subgraph. This metric is distinct from the target-ranking accuracy reported in Table~\ref{tab:ranking-baselines}: the latter evaluates how highly the designated target is ranked among candidate targets, whereas target preservation evaluates whether a given target is retained after graph reduction. Hence, 100\% target preservation does not imply 100\% Top-1 ranking accuracy. In contrast, $k$-hop fails to preserve the target when $k$ is too small to span the initial-to-target distance and requires domain knowledge to select $k$; for the IoT use case ($^{*}$), the smallest $k$ covering all ten scenarios is $k=16$, at which the average $\alpha=0.95$, compared with $0.06$ for the proposed method. The knowledge-equipped forward-only and all-paths baselines (Section~\ref{sec:five}) likewise preserve the target but achieve substantially less reduction.

\subsection{Optimal Path Preservation}

Optimal-path preservation holds in every case: both the minimum-time path and the minimum-cost path in the reduced graph match those in the full graph, and attack trees built from reduced graphs contain the same optimal sequence and minimum traversal time as those from the complete graph, with far fewer nodes and branches to analyze. Importantly, the synthesized tree is not limited to the optimal route: it enumerates all feasible paths in the reduced subgraph as OR-alternatives (with sequential steps represented as sequential-AND nodes), so alternative attack strategies are retained. For complex facility Scenario~1 the synthesized tree presents all four feasible routes to the data center (Fig.~\ref{fig:multiple-Building-tree}), all of which take 9.5 minutes; the reduced subgraph for the IoT \texttt{WebAttack} scenario likewise retains all eight of its feasible routes.

For the single building, the optimal 6.5 minute path through staircases is preserved in the 7-node reduced graph. For the complex facility Scenario 1, the 9.5 minute path through Building C's security mantrap and internal corridor remains intact in the 9-node subgraph. For the IoT \texttt{WebAttack} scenario, the 30.35 minute four-hop path appears in the 66-node reduced attack graph.

\subsection{Comparison with Full Graph Analysis}
To assess the practical effect of graph reduction on attack-tree
synthesis, we compare the proposed target-oriented subgraph with the
corresponding full graph. For Complex Facility Scenario 1, the original
graph contains 42 nodes and 56 edges, whereas the reduced subgraph
contains 9 nodes and 10 edges, corresponding to an 81\% reduction in
graph elements. Both representations preserve the same optimal attack
sequence. The full-graph construction must nevertheless explore
branches that subsequently become infeasible and are removed during
cost propagation, whereas the reduced representation excludes these
branches before AT synthesis.

This comparison demonstrates a reduction in the structural search space
of the AT-generation procedure. It does not, however, constitute a
measurement of end-to-end runtime or memory consumption. In particular,
because AT synthesis enumerates feasible attack paths, its computational
cost depends not only on $|V|$ and $|E|$ but also on the number of feasible
paths.

\subsection{Ranking-Metric Ablation}
\label{sec:ablation}
To isolate the contribution of each ranking metric (degree centrality $C_{\text{D}}$,
closeness centrality $C_{\text{C}}$, and strategic value $S$), we evaluate every
single-metric and leave-one-out configuration on the five complex-facility scenarios,
reporting top-1, top-3, and mean reciprocal rank (Table~\ref{tab:ablation}).

\begin{table}[t]
\centering
\caption{Ranking-metric ablation on the complex facility (five scenarios).}
\label{tab:ablation}
\begin{tabular}{l c c c}
\hline
\textbf{Metric configuration} & \textbf{Top-1} & \textbf{Top-3} & \textbf{MRR} \\
\hline
Degree centrality only        & 0/5 & 1/5 & 0.25 \\
Closeness centrality only     & 0/5 & 0/5 & 0.16 \\
Value only                    & 1/5 & \textbf{4/5} & \textbf{0.48} \\
Degree + Closeness            & 0/5 & 1/5 & 0.23 \\
Degree + Value                & 1/5 & \textbf{4/5} & \textbf{0.48} \\
Closeness + Value             & 1/5 & \textbf{4/5} & \textbf{0.48} \\
All three ($0.2,0.3,0.5$)     & 1/5 & \textbf{4/5} & \textbf{0.48} \\
\hline
\end{tabular}
\end{table}

The ablation shows that \textbf{strategic value is the necessary and dominant metric} at this multi-target facility: every configuration that includes $S$ reaches top-3 $4/5$ and MRR $0.48$, whereas the centrality-only configurations (degree, closeness, degree+closeness) place the target in the top three at most once. The two centrality terms therefore act as secondary discriminators rather than primary signals in the evaluated complex-facility scenarios. We retain all three metrics because closeness and degree can provide additional structural information when strategic values are uncalibrated or tied, as in the single-building case where they help distinguish locations with similar strategic values. The current results, however, do not establish a consistent improvement in ranking performance from combining the centrality terms with strategic value. Given the limited number of scenarios covered in this work, the results should be interpreted as an illustrative proof of concept for the target-ranking procedure rather than as a statistically robust
validation of target-prediction performance.

\subsection{Sensitivity Analysis}
\label{sec:sensitivity}

\subsubsection{Weight Parameter Analysis}

The strategic ranking function depends on weight parameters $w_D$, $w_C$, and $w_S$ that balance degree centrality, closeness centrality, and strategic value. We evaluate the robustness of target identification across different weight configurations to determine whether the approach requires careful parameter tuning.

For the single building scenario, we generate all 66 valid weight combinations on a 0.1-step simplex grid subject to the constraint $w_D + w_C + w_S = 1.0$. For each configuration, we recompute the strategic ranking and identify the top-ranked target. The server room ranks first in 50 of the 66 configurations; the exceptions occur only when strategic value is heavily discounted ($w_S \leq 0.1$), where a more central but less valuable office (office\_1a or office\_2a) overtakes it: the server room ranks first in none of the 11 configurations with $w_S = 0$ and 5 of the 10 with $w_S = 0.1$. Restricting to configurations with $w_S \geq 0.2$, the server room is identified in all 45 cases.

For the complex facility, we perform the same analysis on Scenario 1. The primary data center server room ranks first in 40 of 66 configurations, and in 36 of the 45 configurations with $w_S \geq 0.2$ (80\%). All nine remaining cases occur at $0.2 \le w_S \le 0.4$ and rank the executive office $a\_exec\_1$ first: its strategic value (0.90) is only slightly below that of $c\_server\_2$ (0.95), but its closeness centrality is about twice as high, so it can outrank the data center when $w_S \le 0.4$. For $w_S \ge 0.5$, the data center server room ranks first in all 21 configurations. These cases reflect competition among several high-value assets (the facility contains multiple targets rated $\ge$ 0.8) rather than the selection of a low-value node. This is why we report the ranking as a defender \emph{shortlist} (Section~\ref{sec:ranking-baselines}) rather than a single top-1 prediction.

We compute decision boundaries by determining the minimum weight threshold for stable identification. For the single building, $w_S \ge 0.2$ suffices (45 of 45 configurations), whereas at $w_S = 0.1$ the server room ranks first in 5 of 10. For the complex Scenario 1, stable identification requires $w_S \ge 0.5$ (21 of 21 configurations). With $w_S \ge 0.2$, the data center ranks first in 80\% of configurations (36 of 45). The default weights ($w_S = 0.5$) lie within the stable region of both cases. The higher threshold in the complex facility indicates that strategic value should receive sufficient weight in environments with multiple competitive high-value assets.

\subsubsection{Strategic-Value Sensitivity}
Because the ablation identifies strategic value as the dominant metric, we also evaluate the sensitivity of target identification to perturbations of the strategic-value prior $S(v)$. We jitter every node's strategic value by a uniform $\pm p\%$ (200 trials per level) and record how often the top-ranked target remains unchanged. For the single-building case, the original target is preserved in 84\%, 60\%, and 50\% of trials at $p=10,20,30$, respectively. For complex Scenario~1, the corresponding rates are 58\%, 38\%, and 24\%. These results indicate that target identification is relatively stable to small perturbations when one asset is clearly dominant, but is substantially more sensitive in the multi-target facility, where several assets have similar strategic values. Thus, the analysis indicates robustness to the selection of ranking weights within the identified operating range, but sensitivity to inaccuracies in the underlying strategic-value prior. The latter should therefore be treated as an important source of uncertainty in target inference.

\subsubsection{Target Stability Analysis}

We examine how variations in the elapsed time parameter $\Delta t$ affect target identification and graph reduction. Specifically, we study whether the proposed method will always identify the target and optimal reduced graph, given adequate time that covers the ground truth. For the single building, we vary $\Delta t$ from 4 to 12 minutes. The server room is unreachable below its optimal traversal time (6.5 minutes); once reachable, the time-feasible subgraph is stable, with $\alpha = 0.68$ for all $\Delta t \geq 8$ minutes. Because the target-oriented subgraph $G_{\text{sub}}$ (Section~\ref{subsec:graph-reduction}) retains only nodes on a feasible $v_{\text{init}}\!\to\!v_{\text{target}}$ route within budget, it does not expand merely because $\Delta t$ grows: enlarging the budget admits new nodes only when they lie on a newly feasible route.

For the complex facility Scenario 1, varying $\Delta t$ from 10 to 20 minutes leaves the reduction stable at $\alpha = 0.19$ throughout, because the approach to the deep data-center target follows a single corridor and no additional nodes become path-feasible. This stability is desirable: the reduction is robust to the exact choice of $\Delta t$ within the operational range, yet still excludes the target when the budget falls below its optimal traversal time.

 \section{Conclusion and Future Work}
\label{sec:conclusion}

In this work, we address the challenge of synthesizing ATs in situations where the attacker’s objective is not known in advance. By combining temporal reachability analysis with strategic target ranking, we introduce a systematic way to narrow the search space of potential attack targets. The method focuses on feasible attack paths and on states that can be reached within given temporal constraints. This leads to a reduced graph that supports the construction of concise and interpretable ATs without losing completeness. We implement and evaluate the method in building environments and IoT attack scenarios, with target inference evaluated in the building cases and target-conditioned graph reduction and attack-tree construction evaluated on the CICIoT2023 topology using supplied targets. The reduction factor is $\alpha = 0.68$ on the single building and as low as $0.14$ on the multi-building complex, and averages $\alpha = 0.056$ on the IoT topology; in every case coverage is $100\%$, and the optimal path and all feasible attack paths in the evaluated reduced subgraphs are preserved.

We note a scope limitation regarding target inference: the strategic ranking infers a target only when the environment contains a clear, dominant, strategically distinguishable high-value asset, as in the building cases. It does not transfer to flat, gateway-centric IoT topologies, where centrality is dominated by the gateway and no single asset is distinguishable; the IoT use case therefore demonstrates graph reduction with a supplied target. On the CICIoT2023 topology we found a purely structural criterion insufficient: because nearly every device is reachable through the gateway, the ranking returns the same central hub regardless of the observation point, even after excluding the gateway. A viable route, left to future work, is to condition target ranking on an \emph{attack-type prior} from the dataset's documented per-attack victim profiles; such a prior discriminates only for targeted attack classes (e.g.\ web-based and brute-force, whose victims are specific device types), not indiscriminate floods. Device-level ground truth is itself limited, as the dataset identifies most victims by model rather than by a network identifier matchable to the captured topology.

We also conduct a sensitivity analysis that distinguishes two sources of uncertainty in target inference. The ranking is relatively stable with respect to the selection of weights when the strategic-value weight satisfies the identified thresholds ($w_S \ge 0.2$ in the single building and $w_S \ge 0.5$ in the complex facility). However, perturbations of the strategic-value prior itself produce a marked decrease in target stability, particularly in multi-target environments. These results indicate that the method is less sensitive to moderate weight-selection variations within the identified operating range, but remains sensitive to errors in the defender-supplied strategic-value prior. Consequently, careful calibration of strategic values remains important for reliable target inference. A further limitation concerns the attacker model. The synthesized AT is conditioned on the observation available at a single elapsed time $\Delta t$ and assumes static traversal costs, so it does not account for an adversary that adapts its strategy in response to defender countermeasures, as is characteristic of advanced persistent threats. The framework does support re-invocation as new observations arrive, since observability $\mathcal{O}(v,t)$ is explicitly time-varying and the reduction stage requires only two shortest-path computations, which are almost linear in the size of the graph, whereas the subsequent AT synthesis depends on the number of feasible attack paths. Overall, the framework connects partial intrusion observations to structured threat modeling. Future work may explore integration with real-time monitoring systems for incremental re-inference, game-theoretic modeling of adaptive attacker--defender interaction, extensions toward probabilistic reasoning, and adaptation to dynamic environments where
system states evolve continuously.

\bibliographystyle{IEEEtran}
\bibliography{refs}
\end{document}